\documentclass[journal,transmag]{IEEEtran}
\usepackage[T1]{fontenc}
\usepackage{cite}
\usepackage{supertabular} 
\usepackage{graphicx}
\usepackage{lineno,hyperref}
\usepackage{amsmath}
\usepackage{amsfonts}
\usepackage{amssymb}
\usepackage{array,multirow,makecell}
\modulolinenumbers[5]
\usepackage{lscape}
\begin{document}

\title{A Survey on Privacy-preserving Schemes for Smart Grid Communications}

\author{\IEEEauthorblockN{Mohamed Amine Ferrag\IEEEauthorrefmark{1},
Leandros A. Maglaras\IEEEauthorrefmark{2},~\IEEEmembership{Senior Member,~IEEE},
Helge Janicke\IEEEauthorrefmark{2} and Jianmin Jiang \IEEEauthorrefmark{3}}
\IEEEauthorblockA{\IEEEauthorrefmark{1}Department of Computer Science, Guelma University, Algeria}
\IEEEauthorblockA{\IEEEauthorrefmark{2}School of Computer Science and Informatics, De Montfort University, United Kingdom}
\IEEEauthorblockA{\IEEEauthorrefmark{3}Research Institute for Future Media Computing, Shenzhen University, China}

\thanks{Manuscript received 2016; M. A. Ferrag (email: mohamed.amine.ferrag@gmail.com / Phone: +213 661-873-051) L. A. Maglaras (email: Leandros.maglaras@dmu.ac.uk / Phone: +44 116 2078483) H. Janicke (email: heljanic@dmu.ac.uk / Phone: +44 116 257 7617) Jianmin Jiang (email: jianmin.jiang@szu.edu.cn)}}

\markboth{IEEE}
{Shell \MakeLowercase{\textit{et al.}}: Bare Demo of IEEEtran.cls for IEEE Transactions on Magnetics Journals}
\IEEEtitleabstractindextext{%
\begin{abstract}
In this paper, we present a comprehensive survey of privacy-preserving schemes for Smart Grid communications. Specifically, we select and in-detail examine thirty privacy preserving schemes developed for or applied in the context of Smart Grids. Based on the communication and system models, we classify these schemes that are published between 2013 and 2016, in five categories, including, 1) Smart grid with the advanced metering infrastructure, 2) Data aggregation communications, 3) Smart grid marketing architecture, 4) Smart community of home gateways, and 5) Vehicle-to grid architecture. For each scheme, we survey the  attacks of leaking privacy, countermeasures, and game theoretic approaches. In addition, we review the survey articles published in the recent years that deal with Smart Grids communications, applications, standardization, and security. Based on the current survey, several recommendations for further research are discussed at the end of this paper.
\end{abstract}

}

\onecolumn

\maketitle
\IEEEdisplaynontitleabstractindextext
\IEEEpeerreviewmaketitle

\section{Introduction}
\IEEEPARstart{E}{lectricity} is the energy of the future, which is growing day by day relative to that of gas, which is growing more modestly, and that of oil, which is clearly receding \cite{1}. This growth in electricity consumption is due to the development of new information and communication technologies and a climate imperative, i.e., reducing greenhouse gas emissions. The European Union (EU) is committed to reducing its energy consumption by 20\% (compared to expected levels) by 2020 \cite{159}. However, how can we control the electrical energy? The adaptation of Smart Grid is coming to answer this question. The Smart Grid is defined by the U.S. Department of Energy as an electrical system capable of intelligently integrating the actions of different users, consumers, and / or producers in order to maintain an efficient, sustainable, economical and secure electricity supply \cite{160}.

\begin{figure}[h]
 \centering
 \includegraphics[width=1\linewidth]{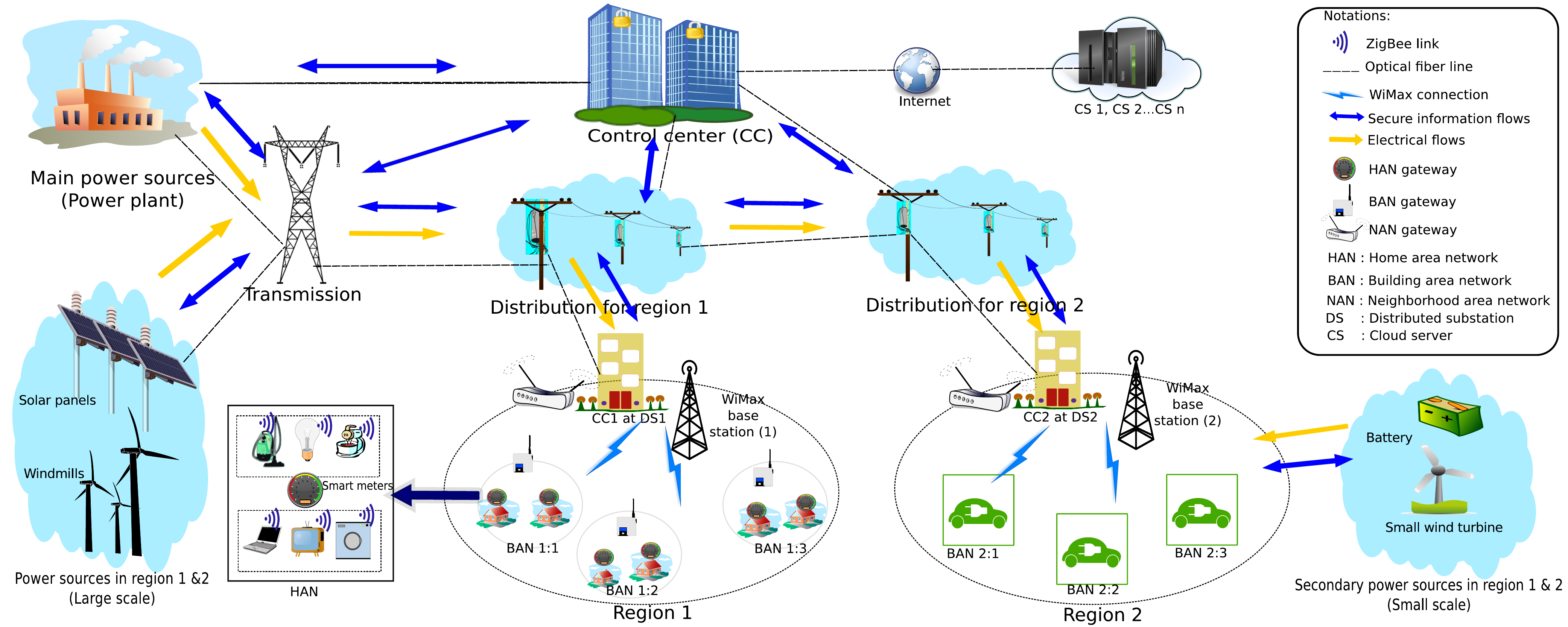}
 \caption{Smart grid architecture}
 \label{fig:Fig1a}
 \end{figure}
 
As shown in Fig. ~\ref{fig:Fig1a}, the Smart grid architecture consists of three types of network architecture (i.e., NAN, BAN, and HAN), including a control center ($CC$) and some cloud servers $CS=\{{CS}_1,{CS}_2,{CS}_3,\ \dots ,\ {CS}_n\}\ $ \cite{1}. The Home Area Network (HAN) uses two types of digital networks, namely, Local Area Network (LAN) and Wide Area Network (WAN). The BAN network connect various departmental networks within a single building, where the rates should be very high. The Neighborhood area network (NAN) provides a secure communication channel between the power company and independent users (a power generator) and the different types management of connected meters (e.g.,water, gas, electric). To transmit and control the energy consumption data between these networks, a gateway can be placed in each type of network.

With the use of control software that will constantly improve power consumption and optimize costs, the future smart grid can improve the security and reliability of the existing power grid \cite{63}. Nevertheless, the Smart Grid cannot be widely deployed without considering the security requirements, namely, authentication, integrity, non-repudiation, access control, and privacy. For this, recently, researchers in the field of computer security have proposed several privacy-preserving schemes for Smart Grid communications. As a result, we are motivated to investigate this schemes and provide a comprehensive and systematic review of the recent studies on published privacy-preserving schemes for Smart Grid communications. More precisely, we select and in-detail examine thirty privacy-preserving schemes that are published between 2013 and 2016. See Tab. \ref{tab:Tab1a} for a breakdown of publication dates.

The main contributions of this paper are:

\begin{itemize}

\item  We provide a classification for the attacks of leaking privacy in Smart Grids, including, key-based attacks, data-based attacks, impersonation-based attacks, and physical-based attacks.

\item  We present various countermeasures and game theoretic approaches used in privacy-preserving schemes for Smart Grids.

\item  We present a side-by-side comparison in a tabular form for the current state-of-the-art of privacy-preserving schemes (thirty) proposed for Smart Grids.

\item  We present a discussion of technical challenges and open directions for future research. In particular, we believe that more work is needed on 1) Attacks such as Sybil attacks, Forgery attacks, and Wormhole attacks, 2) Intrusion Detection mechanisms 3) IoT-driven Smart Grids, 4) Metrics for interdependent privacy, 5) Cloud computing services in Smart Grids,  6) Privacy for Internet of Energy (IoE), and 7) Ethics and Privacy.
\end{itemize}
 
The rest of this paper is organized as follows. In section \ref{sec:survey-articles-for-smart-grids}, we summarize the existing survey works about Smart Grid communications, Smart Grid applications, Smart Grid security, and Smart Grid privacy. In section \ref{sec:threats-and-requirements}, we present an overview of privacy attacks in Smart Grids and defense schemes. Then in section \ref{sec:countermeasures,-game-theoretic-and-formal-proof-approaches}, we discuss the countermeasures, game theoretic and formal proof approaches. In section \ref{sec:privacy-preserving-schemes-for-smart-grids}, we present a side-by-side comparison in a tabular form for the current state-of-the-art of privacy-preserving schemes proposed for Smart Grids. Finally, we identify the future directions and give the conclusion in section \ref{sec:open-questions}, \ref{sec:conclusion}, respectively.

\begin{table}[h]
\centering
\caption{Publication date breakdown - surveyed papers}
\begin{tabular}{||p{0.5in}|p{0.2in}|p{0.5in}|p{0.2in}|p{0.5in}|p{0.2in}||} \hline 
Paper & Year & Paper & Year & Paper & Year \\ \hline 
Sun et al. \cite{49} & 2013 & Li et al. \cite{53} & 2015 & Tsai et al. \cite{108} & 2016 \\ \hline 
Wen et al. \cite{71} & 2013 & Chen et al. \cite{55} & 2015 & Dimitriou et al. \cite{94} & 2016 \\ \hline 
Liang et al. \cite{77} & 2013 & Chen et al. \cite{58} & 2015 & Ni et al. \cite{93} & 2016 \\ \hline 
Wen et al. \cite{70} & 2014 & Bao et al. \cite{62} & 2015 & He et al. \cite{92} & 2016 \\ \hline 
Li et al. \cite{69} & 2014 & Jiang et al. \cite{64} & 2015 & Bao et al. \cite{90} & 2016 \\ \hline 
Fan et al. \cite{76} & 2014 & Abdallah et al. \cite{126} & 2016 & He et al. \cite{96} & 2016 \\ \hline 
Jia et al. \cite{51} & 2014 & Saxena et al. \cite{125} & 2016 & Wang et al. \cite{91} & 2016 \\ \hline 
Saxena et al. \cite{33} & 2015 & Liu et al. \cite{124} & 2016 & Rahman et al. \cite{88} & 2016 \\ \hline 
Shi et al. \cite{35} & 2015 & Han et al. \cite{118} & 2016 & Tan et al. \cite{87} & 2016 \\ \hline 
Deng et al. \cite{36} & 2015 & Wan et al. \cite{89} & 2016 & Gong et al. \cite{83} & 2016 \\ \hline 
\end{tabular}
\label{tab:Tab1a}
\end{table}

\section{Survey articles for Smart Grids}\label{sec:survey-articles-for-smart-grids}
There around forty survey articles published in the recent years that deal with smart grids communications, applications, standardization and security. These survey articles are categorized as shown in tables \ref{tab:Tab1} and \ref{tab:Tab2}.
\begin{table}[h]
\centering
\caption{Areas of research of each survey article for Smart Grids}
\begin{tabular}{||p{1in}|p{0.4in}|p{0.5in}|p{0.6in}|p{0.8in}|p{0.4in}|p{0.5in}|p{0.38in}|p{0.2in}||} \hline 
\textbf{Reference} & \textbf{Security/\newline Privacy} & \textbf{Communi-\newline cations} & \textbf{Applications} & \textbf{Energy management / Power Electronics} & \textbf{Standard-\newline ization} & \textbf{Game Theoretic/ Complex Networks} & \textbf{Projects / demos} & \textbf{V2G} \\ \hline 
\cite{2},\cite{6},\cite{22},\cite{25},\cite{37},\cite{38},\cite{39},\cite{41},\cite{61},\cite{68}, \cite{161}& \checkmark &  &  &  &  &  &  &  \\ \hline 
\cite{3} &  & \checkmark &  &  & \checkmark &  &  &  \\ \hline 
\cite{4},\cite{120} & \checkmark & \checkmark &  &  &  &  &  &  \\ \hline 
\cite{5} & \checkmark & \checkmark &  & \checkmark & \checkmark &  &  &  \\ \hline 
\cite{7},\cite{11},\cite{20},\cite{21},\cite{24},\cite{30},\cite{40} &  & \checkmark &  &  &  &  &  &  \\ \hline 
\cite{8},\cite{123} &  & \checkmark & \checkmark &  &  &  &  &  \\ \hline 
\cite{9},\cite{10},\cite{28} &  &  &  &  &  &  & \checkmark &  \\ \hline 
\cite{12} &  &  &  &  & \checkmark &  &  &  \\ \hline 
\cite{13},\cite{30} &  &  &  & \checkmark &  &  &  &  \\ \hline 
\cite{14},\cite{23} &  &  &  &  &  & \checkmark &  &  \\ \hline 
\cite{15} &  & \checkmark &  &  & \checkmark &  &  &  \\ \hline 
\cite{16},\cite{121} & \checkmark & \checkmark &  & \checkmark &  &  &  &  \\ \hline 
\cite{17},\cite{18} &  & \checkmark &  & \checkmark &  &  &  &  \\ \hline 
\cite{19} & \checkmark & \checkmark &  &  & \checkmark &  &  &  \\ \hline 
\cite{26} &  &  &  & \checkmark &  & \checkmark &  &  \\ \hline 
\cite{27},\cite{29} &  &  & \checkmark &  &  &  &  &  \\ \hline 
\cite{32} &  &  &  &  &  & \checkmark &  &  \\ \hline 
\cite{95} &  & \checkmark & \checkmark & \checkmark & \checkmark &  &  &  \\ \hline 
\cite{117} &  &  &  &  &  &  &  & \checkmark \\ \hline 
\cite{25} & \checkmark &  &  &  &  &  &  & \checkmark \\ \hline 
\cite{122} &  &  &  & \checkmark & v &  &  &  \\ \hline 
\end{tabular}
\label{tab:Tab1}
\end{table}
 
\begin{table}[h]
\centering
\caption{Survey articles for Smart Grids grouped by area of research}
\begin{tabular}{||p{2in}|p{1in}||} \hline 
\textbf{Reference} & \textbf{Area of research} \\ \hline 
\cite{2},\cite{4} \cite{6}, \cite{22}, \cite{25}, \cite{37}, \cite{38}, \cite{39}, \cite{41}, \cite{61}, \cite{68},\cite{16},\cite{19},\cite{120},\cite{121},\cite{161} & Security / Privacy \\ \hline 
\cite{3},\cite{4},\cite{5},\cite{6},\cite{7},\cite{11},\cite{15},\cite{16},\cite{17},\cite{18},\cite{19},\cite{20},\cite{21},\cite{24},\cite{30},\cite{40},\cite{95},\cite{120},\cite{121},\cite{123} & Communications \\ \hline 
\cite{8},\cite{27},\cite{29},\cite{95} &  Applications \\ \hline 
\cite{3},\cite{5},\cite{12},\cite{15},\cite{19},\cite{95},\cite{122} & Standardization \\ \hline 
\cite{14},\cite{23},\cite{26},\cite{12} & Game theoretic approaches / Complex Networks \\ \hline 
\cite{9},\cite{10},\cite{28} & Projects \\ \hline 
\cite{5},\cite{13},\cite{16},\cite{17},\cite{18},\cite{26},\cite{31},\cite{95},\cite{121},\cite{122} & Energy Management /         Power Electronics \\ \hline 
\cite{25}, \cite{117} & Vehicle to Grid Technology \\ \hline 
\end{tabular}
\label{tab:Tab2}
\end{table}
\begin{table}[h]
\centering
\caption{Year of publication}
\begin{tabular}{||p{2.2in}|p{0.4in}||} \hline 
\textbf{Reference} & \textbf{ Year} \\ \hline 
\cite{10},\cite{12},\cite{20},\cite{21} & 2010 \\ \hline 
\cite{3},\cite{9},\cite{14},\cite{15},\cite{24},\cite{31} & 2011 \\ \hline 
\cite{2},\cite{37},\cite{4},\cite{11},\cite{18},\cite{23},\cite{27},\cite{28},\cite{30},\cite{32} & 2012 \\ \hline 
\cite{6},\cite{61},\cite{5},\cite{7},\cite{8},\cite{13},\cite{16},\cite{17} & 2013 \\ \hline 
\cite{22},\cite{38},\cite{19},\cite{26},\cite{29} & 2014 \\ \hline 
\cite{39}, \cite{41},\cite{117},\cite{120} & 2015 \\ \hline 
\cite{25}, \cite{68},\cite{40},\cite{95},\cite{121},\cite{122},\cite{123} \cite{161} & 2016 \\ \hline 
\end{tabular}
\label{tab:Tab3}
\end{table}

\subsection{Smart Grid Communications}
Several articles describe the communication requirements and capabilities that can be used or combined in order to support the smart behavior of a grid \cite{3,4,5,6,7}, \cite{11}, \cite{15,16,17,18,19,20,21}, \cite{24}, \cite{30}, \cite{40}, \cite{95},\cite{120,121}. Authors in \cite{19} focus on wireless communication networking technologies for smart grid neighborhood area networks (NANs). Authors in \cite{20} survey different opportunities and challenges of applying WSNs in smart and present several field tests that have been performed on IEEE 802.15.4-compliant wireless sensor nodes in real-world power delivery and distribution systems. These field trials were performed in order to measure background noise, channel characteristics, and attenuation in the 2.4-GHz frequency band. Gomez et al, in \cite{21} survey the most relevant current and emerging solutions suitable for Wireless home automation networks at Grid: ZigBee, Z-Wave, INSTEON, Wavenis, and IP-based solutions. Following a different approach, authors in \cite{16} and \cite{30} specifically focus on how cloud computing (CC) could be used for energy management, information management and security of the Smart Grid. In two similar survey works  \cite{24},\cite{40} that were published in 2011 and 2016 respectively, authors research how cognitive radio would support smart grid idea including system architecture, communication network compositions, applications, and CR-based communication technologies.

\subsection{Smart Grid Applications}
Several survey articles \cite{8},\cite{27},\cite{29},\cite{95} focus on the applications that can be supported from a Smart Grid, while on the works presented in \cite{3},\cite{5},\cite{12},\cite{15},\cite{19},\cite{95} authors discuss standardization issues that are related to Smart Grids. Energy management and Power electronics are two important aspects of a smart grid and these issues are surveyed in \cite{5},\cite{13},\cite{16},\cite{17},\cite{18},\cite{26},\cite{31},\cite{95}. In \cite{13} authors discuss demand response potentials and benefits in smart grids. The potential of using game theory-based methods to solve different problems in smart grid are surveyed in \cite{14},\cite{23},\cite{26},\cite{12}. Especially in \cite{12} authors present a very interesting survey article for Smart Grids. The article explores the most important scientific studies that used Complex Network Analysis techniques and methodologies for investigating the properties of several Power Grids infrastructures. Tan et al \cite{117}, review the concept, framework, advantages, challenges and optimization strategies of V2G technology, which is one of the smart grid technologies that involves the Electric Vehicle (EV) to improve the power system operation. Authors in \cite{123} discuss about the electric grid as a composition of sub-systems with increasing autonomy that interact in order to transform the grid into a smart system, forming a system-of-systems. Since the current survey article focuses on privacy preserving techniques on Smart Grids, we present in the next subsection a detailed analysis of the published survey articles that focused on smart grid security.

\subsection{Smart Grid Security}
Liu et al. \cite{37} presented back in 2012 an early survey about cyber security and privacy issues in the Smart Grid. The article concludes that privacy in Smart Grids may be addressed by adopting newly anonymous and camouflage communication technologies. The same year, authors in \cite{2} discuss specific security requirements that a smart grid has, along with challenges and current solutions. They describe several existing solutions that have been implemented or tested on real industry environments that are related to privacy protection, integrity, authentication and trusted computing. Regarding privacy, authors describe briefly seven encryption and anonymization techniques and do not provide any critical comparison of the presented methods. Also since the report was published on 2012 it doesn't cover novel methods that were introduced the previous 5 years. One year later in 2013, Wang et al. in \cite{6} provides a detailed survey of cyber security issues and challenges for Smart Grids.  Similar to \cite{2}, authors present current requirements that a Smart Grid has making it a demanding environment in terms of both security and reliability.  Authors also categorize and evaluate network threats using case studies and describe network and cryptographic countermeasures against cyber attacks.  By devoting a section for privacy, authors present and analyze in depth cryptographic, authentication and key management, along with case studies for every sub-category. The article concludes that a tradeoff between latency and privacy is a major issue in Smart Gird security, where physical layer authentication can be a solution especially for wireless communications. Key management techniques can also increase the privacy that is reassured from the system and the combination of different communication capabilities can help increase system's performance as long as secure communication is reassured. The survey is very detailed and analytical but since 2013 new advanced security methods and privacy solutions have been proposed that the current article tries to critically discuss and analyze. In 2014, Komninos et al. \cite{22}, discusses open issues, challenges and countermeasures for smart grid and smart home security. This article, compared to the previous ones, emphasizes mostly on smart home security and how this is combined with the concept of Smart Grid security. Authors present major security goals that are expected to be met and identify threats that may occur under representative scenarios of interaction between Smart Home and Smart Grid entities along with the impact that they have on the system. Authors also describe several privacy preserving methods that are based on anonymization, encryption, perturbation, verifiable computation models and obfuscation. The main conclusion of the article regarding privacy is the need of a legal framework specific to privacy in the Smart Grid, the establishment of new key management techniques and new aggregation mechanisms. The article doesn't focus on privacy preserving schemes, as the current survey does, but discuss the general concept of smart grid security viewed under the smart home interaction perspective.  Han et al. \cite{25}, in 2016 presents an analysis of privacy preservation issues for V2G networks. Authors summarize current solved problems, various techniques used and their pros and cons but their work is limited only for V2G communications that is a small part of a Smart Grid network.  Song et al. \cite{68}, in 2016 presents another survey article that discusses security advances of Smart Grid from a data driven approach. On this aspect Song thoroughly investigates data generation, data acquisition, data storage, data processing and data analytics aspects of smart grid security.  The article presents future open issues, such as security of plug in electric vehicles, transactive energy and architectures and frameworks in context of Internet of Things, but doesn't focus on privacy preserving schemes that the current survey does. In addition, Cintuglu et al. \cite{161} present a survey on Smart Grid cyber-physical system testbeds by providing a four step taxonomy based on Smart Grid domains, research goals, test platforms, and communication infrastructure.

\subsection{Smart Grid Privacy}
There exist four surveys that are recently published and are focused on privacy preserving for Smart Grids \cite{38,39,41,61}. Authors in \cite{61} present the state of the art of privacy preserving techniques, focusing mostly on data aggregation. The article that was published in 2013 showcases the existence of sophisticated methods that can disaggregate the measurements and provide an accurate estimation of the moment when each appliance is turned on and off.  Due to this, authors decided to focus on data aggregation techniques that are used to preserve privacy in a Smart Grid. Based on their analysis they define three main categories of challenges, those related to secure cryptographic protocols, those related to hardware limitations and finally those related to signal processing. Authors back in 2014 \cite{38} presented state of the art approaches that are related to the problem of customer privacy-protection in the smart grid.  In order to so, they introduce two terms that describe important problems that have to be solved concerning privacy and smart metering: metering for billing and metering for operations. They argue that there is a tradeoff between sampling frequency attribution and exactness on the one hand and privacy on the other hand.  Accurate and frequent metering can give the necessary information to the billing and management company of a Smart Grid but imposes threats regarding privacy of the consumers. They categorize the different methods in different groups, e.g. aggregation, cryptography, anonymization, imprecise data etc. Uludag et al. \cite{39}, in a chapter that was published in 2015 focuses on the problem of determining sophisticated usage patterns from the smart meter data and the countermeasures. After trying to define the term privacy by viewing it from different angles, e.g. personal, information, organization and intellectual, Uludag tries to define the different privacy concerns that are related to a Smart Grid. By performing a detailed taxonomy of solutions that were published until 2015, the article identifies the various strengths and weaknesses of each method and discusses open issues and future opportunities. Finally, the same year (2015) de Oliveira \cite{41} publishes a report detailing Privacy-Preserving Protocols for Smart Metering Systems. 

All of the aforementioned surveys that are related to privacy preservation for Smart Grids are published until 2015 and don't cover state of the art methods that were recently introduced. The current survey  focuses on privacy preserving methods for Smart Grids and discusses performance and limitations of each scheme. It also presents a detailed taxonomy of the state of the art of privacy preserving methods for smart grids and a full list of all current research efforts and trends. Based on this thorough analysis open issues and future directions are identified that combine both innovative research along with the application, through appropriate adaptation, of existing solutions from other fields. We believe that this study will help researchers focus on the important aspects of privacy issues in the smart grid area and will guide them towards their future research.

\section{Threats and requirements}\label{sec:threats-and-requirements}

\begin{table}[h]
\centering
\caption{Summary of privacy attacks in Smart Grids and defense schemes}
\checkmark indicates fully supported; x indicates not supported; 0 indicates partially supported.
{\tiny\begin{tabular}{||p{0.54in}|p{0.06in}|p{0.06in}|p{0.06in}|p{0.06in}|p{0.06in}|p{0.06in}|p{0.06in}|p{0.06in}|p{0.06in}|p{0.06in}|p{0.06in}|p{0.06in}|p{0.06in}|p{0.06in}|p{0.06in}|p{0.06in}|p{0.06in}|p{0.06in}|p{0.06in}|p{0.06in}|p{0.06in}|p{0.06in}|p{0.06in}|p{0.06in}|p{0.06in}|p{0.06in}|p{0.06in}|p{0.06in}|p{0.06in}||} \hline 
\textbf{} & \multicolumn{29}{|p{5in}||}{\begin{center}
\textbf{Privacy-preserving schemes for Smart Grids}
\end{center}} \\ \hline 
\textbf{Adversary model} & \textbf{\cite{33}} & \textbf{\cite{35}} & \textbf{\cite{51}} & \textbf{\cite{53}} & \textbf{\cite{55}} & \textbf{\cite{58}} & \textbf{\cite{62}} & \textbf{\cite{64}} & \textbf{\cite{36} } & \textbf{\cite{49}} & \textbf{\cite{69}} & \textbf{\cite{70}} & \textbf{\cite{71}} & \textbf{\cite{77}} & \textbf{\cite{76}} & \textbf{\cite{87}} & \textbf{\cite{88}} & \textbf{\cite{89}} & \textbf{\cite{90}} & \textbf{\cite{91}} & \textbf{\cite{92}} & \textbf{\cite{93}} & \textbf{\cite{94}} & \textbf{\cite{96}} & \textbf{\cite{108}} & \textbf{\cite{118}} & \textbf{\cite{124}} & \textbf{\cite{125}} & \textbf{\cite{126}} \\ \hline 
 Key-based attacks\newline \textbf{} & \checkmark & X & X & X\textbf{} & X\textbf{} & X\textbf{} & X\textbf{} & X\textbf{} & X\textbf{} & X\textbf{} & X\textbf{} & X\textbf{} & 0\textbf{} & 0\textbf{} & X\textbf{} & X\textbf{} & X\textbf{} & X\textbf{} & X\textbf{} & X\textbf{} & 0\textbf{} & X\textbf{} & X\textbf{} & 0\textbf{} & \checkmark\textbf{} & X\textbf{} & X\textbf{} & \checkmark & X\textbf{} \\ \hline 
Data-based attacks\newline \textbf{} & \checkmark & \checkmark & \checkmark & \checkmark\textbf{} & X\textbf{} & X\textbf{} & \checkmark\textbf{} & X\textbf{} & 0\textbf{} & 0\textbf{} & 0\textbf{} & 0\textbf{} & X\textbf{} & X\textbf{} & \checkmark\textbf{} & \checkmark\textbf{} & \checkmark\textbf{} & 0\textbf{} & 0\textbf{} & \checkmark\textbf{} & \checkmark\textbf{} & \checkmark\textbf{} & \checkmark\textbf{} & \checkmark\textbf{} & X\textbf{} & X\textbf{} & 0\textbf{} & \checkmark & 0\textbf{} \\ \hline 
Impersonation-based attacks\textbf{} & \checkmark & 0 & 0 & 0\textbf{} & X\textbf{} & X\textbf{} & X\textbf{} & X\textbf{} & X\textbf{} & X\textbf{} & X\textbf{} & X\textbf{} & X\textbf{} & X\textbf{} & X\textbf{} & X\textbf{} & X\textbf{} & \checkmark\textbf{} & \checkmark\textbf{} & 0\textbf{} & \checkmark\textbf{} & 0\textbf{} & 0\textbf{} & \checkmark\textbf{} & \checkmark\textbf{} & \checkmark\textbf{} & 0\textbf{} & \checkmark & \checkmark\textbf{} \\ \hline 
Physical-based attacks\textbf{} & \checkmark & X & X & X\textbf{} & \checkmark\textbf{} & \checkmark\textbf{} & \checkmark\textbf{} & \checkmark\textbf{} & X\textbf{} & X\textbf{} & X\textbf{} & X\textbf{} & \checkmark\textbf{} & \checkmark\textbf{} & X\textbf{} & X\textbf{} & X\textbf{} & X\textbf{} & X\textbf{} & X\textbf{} & 0\textbf{} & X\textbf{} & X\textbf{} & 0\textbf{} & 0\textbf{} & 0\textbf{} & X\textbf{} & \checkmark & X\textbf{} \\ \hline 
\end{tabular}}
\label{tab:Tab4}
\end{table}
 \begin{figure}[h]
 \centering
 \includegraphics[width=1.05\linewidth]{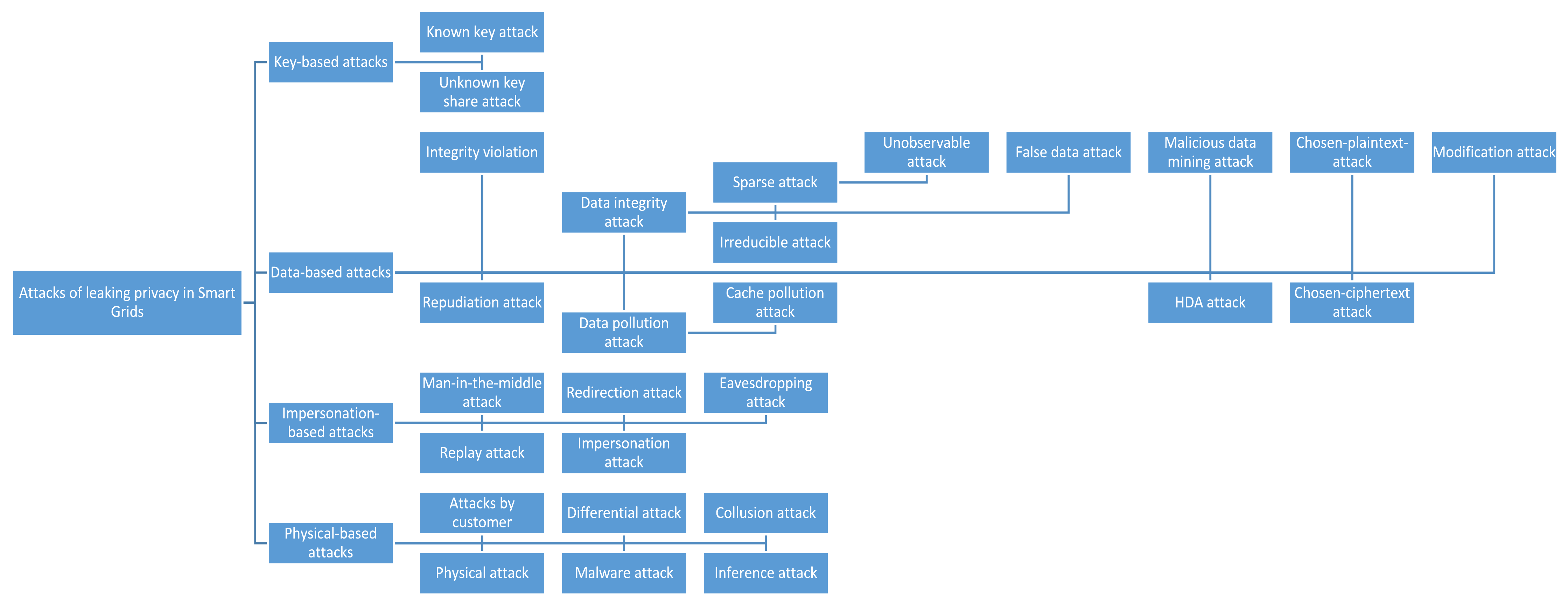}
 \caption{Classification of attacks of leaking privacy in Smart Grids}
 \label{fig:Fig1}
 \end{figure}
 
In this section, we discuss the attacks of leaking privacy in Smart Grids. Generally, the classification of attacks in Smart Grids frequently mentioned in the literature is done using different criteria such as passive or active, internal or external etc. In our survey article, we classify the attacks of leaking privacy in Smart Grids in four categories as shown in Fig. \ref{fig:Fig1}, including, 1) key-based attacks, 2) data-based attacks, 3) impersonation-based attacks, and 4) physical-based attacks. In addition, Tab. \ref{tab:Tab4} provides a detailed description of different privacy attacks in Smart Grids and defense schemes. 
 
\subsection{Attacks of leaking privacy in Smart Grids}
 
\subsubsection{Key-based attacks}
Generally, after the registration phase, each part in a Smart Grid has a secret key or certificate to carry out the authentication phase. At any moment, an adversary (the customer or the operators/ maintenance personnel) can launch both known key attack \cite{144} and unknown key share attack \cite{143} on an authenticated key agreement or authenticated key agreement with key confirmation. More precisely, in known key attack, an adversary records the frequencies in each byte of the plaintexts and ciphertexts. Nevertheless, in the unknown key share attack, an adversary manipulates two entities $A$ and $B\ $where $A$ ends up believing she shares a key with $B$, and although this is in fact the case, $B$ mistakenly believes the key is instead shared with an entity $E\ne A$. Since the private key is different and the public identity of each user is newly generated for each session, both the schemes \cite{33} \cite{125} can prevent the known key attack and the scheme \cite{108} can resist to the unknown key share attack. 
 
  \begin{figure}[h]
  \centering
  \includegraphics[width=0.7\linewidth]{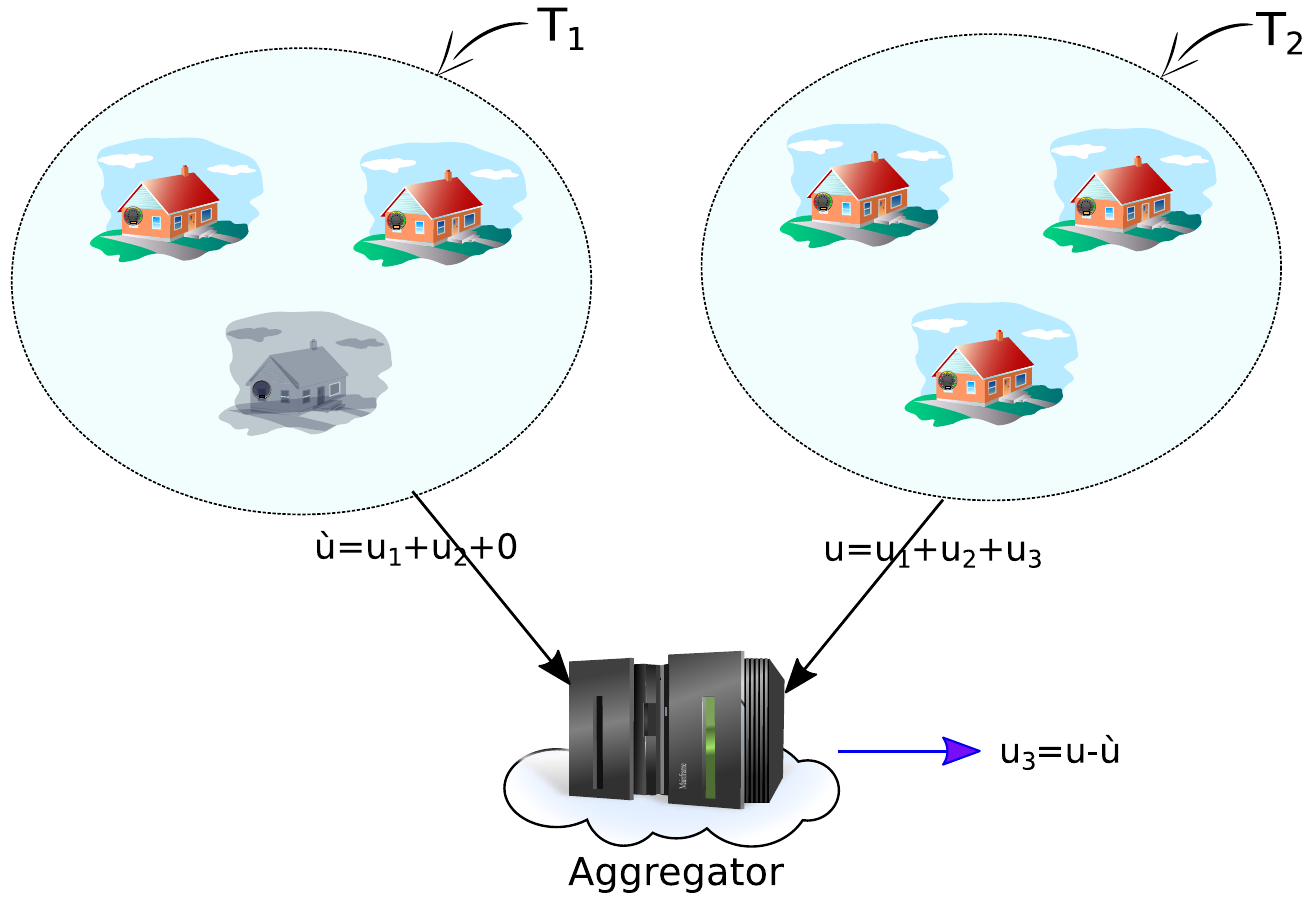}
  \caption{HDA attack in Smart Grid defined by Jia et al. in \cite{51}}
  \label{fig:Fig2}
  \end{figure}
   \begin{figure}[h]
    \centering
    \includegraphics[width=0.7\linewidth]{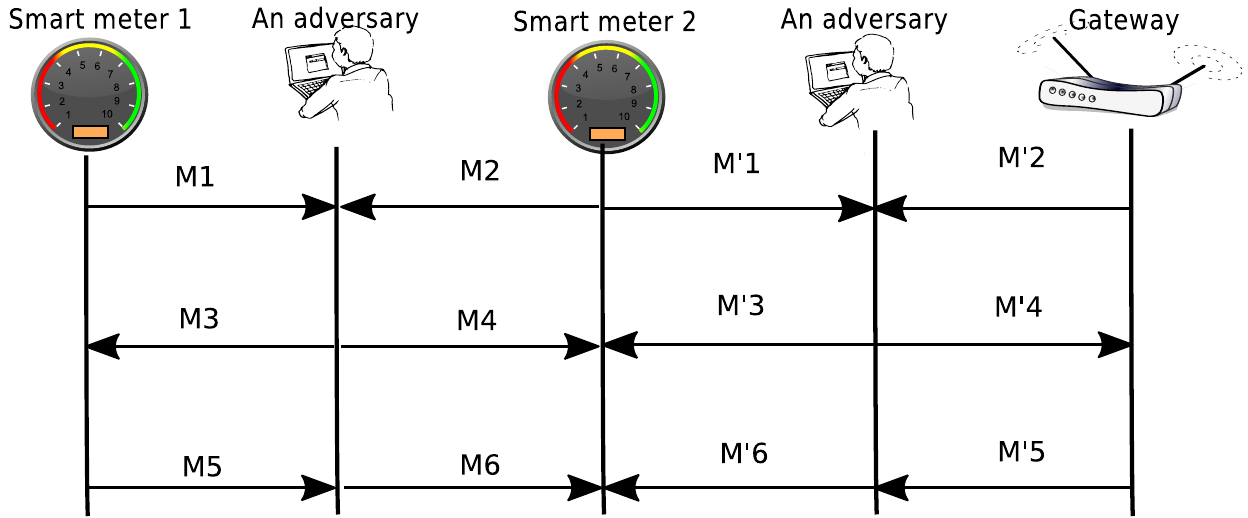}
    \caption{MITM attack defined by Conti et al. in \cite{44}}
    \label{fig:Fig3}
    \end{figure}
\subsubsection{Data-based attacks}
 
To ensure the electricity data integrity in Smart Grids, customers can help by shifting their electricity consumption to different times, which is not always possible as discussed in \cite{145}. However, an adversary can launch several modification operations on the electricity data that contain the information about the electricity consumption. We classify nine attacks in this category of data-based attacks, namely, integrity violation, repudiation attack, malicious data mining attack, human-factor-aware differential aggregation (HDA) attack, chosen-plaintext-attack, chosen-ciphertext attack, data integrity attack, data pollution attack, and modification attack. If an external adversary is successful at modifying the transmitted data, the integrity violation (data loss) can be performed, which is prevented by the scheme \cite{33} using the idea of access control. The repudiation attack \cite{147} refers to a denial of participation in all or part of the communication in Smart Grid, which can be prevented by the scheme \cite{125} using the identity and hash-signature verification. If an adversary is successful at corrupting the aggregator and collaborates with some compromised meters, the malicious data mining can be performed, which can prevented by the scheme \cite{51} based on the privacy-preserving aggregation phase. As defined by Jia et al. \cite{51}, the HDA attack is based on the effect of the human factor on the data aggregate where an adversary can obtain smart meter id3's readings by comparing two aggregation results as shown in Fig. \ref{fig:Fig2}. The scheme in \cite{51} can resist to HDA attack under the Byzantine attack. If an adversary can distinguish a ciphertext when two corresponding plaintexts are given, the chosen-plaintext-attack \cite{149} can be launched, which can be prevented by the scheme \cite{118} using the random layer and the homomorphic layer. If an adversary knows the ciphertexts that are used into the Smart Grid and can obtain the resulting plaintexts, the chosen-ciphertext attack (CCA) \cite{150} can be launched which can lead to the recovery of the hidden secret key used for decryption. The CCA attack can be prevented by the scheme presented in \cite{76} by using an identity-based encryption.
 \begin{table}[h]
 \centering
 \caption{Approaches for detecting and avoiding the MITM attack}
\begin{tabular}{||p{0.6in}|p{1.8in}|p{1.8in}||} \hline 
\textbf{Reference} & \textbf{Data attacked} & \textbf{Approach} \\ \hline 
Saxena et al. (2016) \cite{33} & Message information over the network & Two-factor authentication based on a random private key \\ \hline 
He et al. (2016) \cite{92} & Information between the smart meter and the server provide & Mutual authentication between the smart meter and the server provider based on the lightweight anonymous key distribution \\ \hline 
He et al. (2016) \cite{96} & Information between the aggregator and the user & Authentication between the aggregator and the user using the id-based signature scheme \\ \hline 
Tsai et al. (2016) \cite{108} & Information between the smart meter and the server provide & Authentication between the smart meter and the server provide using the identity-based cryptosystems \\ \hline 
Saxena et al. (2016) \cite{125} & Information between electric vehicles, local aggregator, and certification/registration authority & Authentication based on the dynamic accumulator \\ \hline 
\end{tabular}
\label{tab:Tab5}
\end{table}
 \begin{table}[h]
 \centering
 \caption{Approaches for detecting and avoiding the replay attack}
\begin{tabular}{||p{0.6in}|p{1.7in}|p{0.7in}||} \hline 
\textbf{Reference} & \textbf{Data attacked} & \textbf{Approach} \\ \hline 
Saxena et al. (2016) \cite{33}  & Transmitted message to the user or the server over the network & Timestamp \\ \hline 
Bao et al. (2016) \cite{90}  & Information between the gateway and the cluster head & Timestamp \\ \hline 
He et al. (2016) \cite{92}\newline  & Information between the smart meter and the server provide & Timestamp \\ \hline 
He et al. (2016) \cite{96} & Information between the user and the aggregator & Timestamp \\ \hline 
Han et al. (2016) \cite{118} & Information between battery vehicles and local aggregators & Timestamp \\ \hline 
Saxena et al. (2016) \cite{125} & Information between electric vehicles and local aggregator & Timestamp \\ \hline 
Abdallah et al. (2016) \cite{126} & Information between local aggregators, electric vehicles, and charging stations & Timestamp  \\ \hline 
\end{tabular}
\label{tab:Tab6}
\end{table}
 \begin{table}[h]
 \centering
 \caption{Approaches for detecting and avoiding the impersonation attack}
\begin{tabular}{||p{0.6in}|p{1.7in}|p{1.8in}||} \hline 
\textbf{Reference} & \textbf{Data attacked} & \textbf{Approach} \\ \hline 
Saxena et al. (2016) \cite{33} & Impersonate the users involved in the Smart Grid system (Case-1: Impersonates the maintenance personnel;\newline Case-2: Impersonates the intelligent electronic devices) & Different key pair at each device to prevent the use of old parameter values in other devices \\ \hline 
He et al. (2016) \cite{92}  & Impersonates the smart meter and the server provider & Checking the legality of received messages based on the elliptic curve cryptography \\ \hline 
He et al. (2016) \cite{96}  & Impersonates the users and the aggregator & Checking the legality of received messages based on the id-based signature scheme \\ \hline 
Han et al. (2016) \cite{118} & Impersonates the battery vehicles and the local aggregator & Based on an access authority \\ \hline 
Saxena et al. (2016) \cite{125} & Impersonates the electric vehicles & Mutual authentication between all electric vehicles with local aggregator and/or certification/registration authority \\ \hline 
\end{tabular}
 \label{tab:Tab7}
 \end{table}
  \begin{table}[h]
  \centering
  \caption{Approaches for detecting and avoiding the false data attack and modification attack}
\begin{tabular}{||p{0.6in}|p{1.7in}|p{2in}||} \hline 
\textbf{Reference} & \textbf{Data attacked} & \textbf{Approach} \\ \hline 
He et al. (2016) \cite{92} & The data exchanged between Smart meter and Server provider & By checking the legality of the message authentication code and the digital signature \\ \hline 
He et al. (2016) \cite{96} & The power usage information between a user and an aggregator & The aggregator could find any modification of the message by checking whether the equation $e\left({\sigma }_i,g_1\right)=e(R_iX^{h_i}_i,T)$ \\ \hline 
Ni et al. (2016) \cite{93} & The false consumption data to mislead the control center to make irrational decisions & By using the Schnorr signature to ensure the integrity of the reports during transmission \\ \hline 
Tan et al. (2016) \cite{87} & Retrieve personal consumption information by intercepting the communication between the smart meter and the data aggregator & By using a simple message authentication \\ \hline 
Rahman et al. (2016) \cite{88} & The communication channels between a bidder and a bidding manager or a bidder and a registration manager & By using El-Gamal public key encryption and Schnorr signature scheme \\ \hline 
\end{tabular}
   \label{tab:Tab7a}
   \end{table}
The data integrity attack in Smart Grid consists of a set of compromised power meters whose readings are altered by an adversary \cite{67,151,152,153}. Giani et al. in \cite{67,151} defines some new data integrity attacks such as the unobservable attack and the irreducible attack. Xie et al. in \cite{152} study the economic impact of a data integrity attack in power market operations. Sridhar et al. in \cite{153} is modeling the integrity attacks based on both the Min attack and the Max attack model. Here we must note that we did not find any privacy-preserving scheme that studies the data integrity attack in Smart Grid, but we have found five privacy-preserving schemes \cite{92,96,93,87,88} that can detect both false data attack and modification attack by checking the legality of the message authentication code and the digital signature, as presented in Tab \ref{tab:Tab7a}. As discussed in \cite{154}, with the data pollution attack, an adversary can force the gateways to cache non-popular content in order to affect overall network performance and increase link utilization. Both schemes \cite{93,94} can solve the data pollution attack using orthogonal techniques such as zero knowledge proof.
 
\subsubsection{Impersonation-based attacks}
 
At any time, an adversary can capture data from the Smart Grid transmitted by other smart meters and read the data content in order to recover the information about the energy that is consumed at the smart home. We classify five attacks in this category of impersonation-based attacks, namely, man-in-the-middle attack, replay attack, redirection attack, impersonation attack, and eavesdropping attack. The Man-In-The-Middle (MITM) attack is one of the popular attacks in the new generation networks such as the Smart Grid. As shown in Fig. \ref{fig:Fig3}, an MITM attack is the interception of data passing between two smart meters or a smart meter and a gateway in order to modify the data that pass through, without the victims being aware of it \cite{44}. Specifically, smart meters and the gateway try to initialize secure communication by sending each other their public keys (messages M1, M2, M'1, M'2). An adversary intercept M1,M2,M'1, and M'2, and as a return sends its public key to the victims (messages M3, M4, M'3, M'4). After that, smart meter 1 and gateway encrypts its message by adversary public key, and sends it to smart meter 2 (messages M5 and M'5). Adversary intercepts M5 and M'5, and decrypts it using known private key. Then, adversary encrypts plaintext by smart meter 2 public key, and sends it to smart meter 2 (messages M6 and M'6). We have found five privacy-preserving schemes that can detect and prevent the MITM attack, as presented in Tab. \ref{tab:Tab5}. 
 
The replay attack is a type of MITM attack where an adversary intercept data packets in Smart Grid and replay them to the destination server. We have found six privacy-preserving schemes that can detect and prevent the replay attack, as presented in Tab. \ref{tab:Tab6}, which all these schemes use the timestamp approach. In the V2G networks, an adversary can redirect the vehicles messages to another network out of the original network when accessed. Note that a redirection attack is usually integrated with a phishing attack \cite{155}. The scheme in \cite{125} can defeat redirection attacks using the location verification of each EV by matching received information from the EV with the stored information. During the authentication between the smart meters and gateways, an adversary can initiate an impersonation attack, when he can fake the identity of one of the legitimate parties in the Smart Grid \cite{156}. We have found five privacy-preserving schemes that can detect and prevent the impersonation attack, as presented in Tab. \ref{tab:Tab7}. The eavesdropping attack can occur when an adversary can access the data path of Smart Grid and then can monitor and read all traffic between the smart meters and getaways in order to compromise the privacy of residential users. Based on the Boneh-Goh-Nissim cryptosystem \cite{60} and differential privacy, the scheme \cite{62} is secure against the eavesdropping attack. In addition, the scheme \cite{89} is secure and can detect the eavesdropping attack using an anonymous signature scheme.
 
\subsubsection{Physical-based attacks}
 
An adversary may target the hardware of a battery vehicle, a local aggregator, a gateway or the proxy server to execute an attack such as differential attack, malware attack, collusion attack, and inference attack. To acquire the individual user's data in a Smart Grid, an adversary can lunch a differential attack. The scheme \cite{58} use the differential privacy proposed in the work \cite{59} in order to detect the differential attack. The scheme presented in \cite{62} is secure against differential attack using an appropriate Laplace noise in the form of ciphertext in order to achieve the differential privacy. An adversary can launch a malware attack by deploying undetectable malwares in Smart Grid targeting the privacy disclosure of residential users. Based on the Boneh-Goh-Nissim cryptosystem \cite{60} and differential privacy, the scheme in \cite{62} can resist against the malware attack. By introducing secure self-healing mechanism, the scheme in \cite{64} can resist collusion attack. The scheme in \cite{71} can prevent the cloud server in collusion by using an identity-based encryption scheme in the data encryption phase to encrypt messages. Liang et al. in \cite{77} authors considered collusion attacks launched by multiple compromised homes. Specifically, the scheme in \cite{77} can resist collusion attacks using the proximity score calculation algorithm. If an external adversary is successful at analyzing attributes in database management system, the inference attack can be performed, which is prevented by the scheme presented in \cite{118} using the idea of separable key-chaining management. In order to detect opportunistic attacks in Smart Grid cyber-physical system, li et al. in \cite{172} proposes a scheme using a dirichlet-based probabilistic model \cite{173} that is  used to assess the reputation levels of decentralized local agents.
 
\subsection{Security Requirements}
 
In order to protect Smart Grid communications against the threats mentioned above, privacy-preserving schemes proposed for Smart Grids should satisfy the following security requirements \cite{2,4,6,22,25,37,38,39,41,61,162,68,16,19,120,121}: 
 
\begin{itemize}
 \item  Authentication: When the parts of Smart Grid (smart meters, gateways, local aggregators) want to access in the system, they must initially perform an identification and authentication procedure. The identification is a phase of establishing the identity of the parts. The authentication is a phase that allows the parts to provide proof of identity. 
 
 \item  Integrity: Integrity is the ability to assure that messages exchanged (sent, received or stored) between the parts of Smart Grid has not been modified or deleted. 
 
 \item  Non-repudiation: Non-repudiation in Smart Grid is the ability to prevent a smart meter from denying his involvement in an action in which he participated, for example, denying that the energy reports have been sent by itself.
 
 \item  Access control: Access control is the ability to check whether smart meters, gateways and local aggregators requesting access to a resource have the necessary rights to do it.
 
 \item  Privacy: Privacy of the users in Smart Grid is the ability to protect private information (i.e., identity, location, data aggregation...etc) of smart meters, gateways, and local aggregators.
 \end{itemize}
 
\begin{table}[h]
\centering
\caption{Cryptographic methods used in privacy-preserving schemes for Smart Grids}
\checkmark indicates that the scheme uses the cryptographic method.
{\tiny
\begin{tabular}{||p{0.78in}|p{0.04in}|p{0.04in}|p{0.04in}|p{0.04in}|p{0.04in}|p{0.04in}|p{0.04in}|p{0.04in}|p{0.04in}|p{0.04in}|p{0.04in}|p{0.04in}|p{0.04in}|p{0.04in}|p{0.04in}|p{0.04in}|p{0.04in}|p{0.04in}|p{0.04in}|p{0.04in}|p{0.04in}|p{0.04in}|p{0.04in}|p{0.04in}|p{0.04in}|p{0.04in}|p{0.04in}|p{0.04in}||} \hline 
\textbf{} & \multicolumn{28}{|p{5in}|}{\begin{center}
\textbf{Privacy-preserving schemes for Smart Grids}
\end{center}} \\ \hline 
\textbf{Cryptographic methods} & \textbf{\cite{33}} & \textbf{\cite{35}} & \textbf{\cite{49}} & \textbf{\cite{53}} & \textbf{\cite{55}} & \textbf{\cite{58}} & \textbf{\cite{62}} & \textbf{\cite{64}} & \textbf{\cite{69}} & \textbf{\cite{70}} & \textbf{\cite{71}} & \textbf{\cite{76}} & \textbf{\cite{77}} & \textbf{\cite{83}} & \textbf{\cite{87}} & \textbf{\cite{88}} & \textbf{\cite{89}} & \textbf{\cite{90}} & \textbf{\cite{91}} & \textbf{\cite{92}} & \textbf{\cite{93}} & \textbf{\cite{94}} & \textbf{\cite{96}} & \textbf{\cite{108}} & \textbf{\cite{118}} & \textbf{\cite{124}} & \textbf{\cite{125}} & \textbf{\cite{126}} \\ \hline 
Secure cryptographic hash function \cite{157} & \checkmark\textbf{} & \checkmark\textbf{} & \checkmark\textbf{} & \textbf{} & \checkmark\textbf{} & \checkmark\textbf{} & \textbf{} & \checkmark\textbf{} & \checkmark\textbf{} & \checkmark &  & \checkmark & \checkmark & \checkmark & \checkmark & \checkmark & \checkmark & \checkmark &  & \checkmark & \checkmark & \checkmark & \checkmark &  &  & \checkmark & \checkmark & \checkmark \\ \hline 
PASSERINE public key cryptosystem \cite{134} &  &  &  & \textbf{} &  &  & \textbf{} & \textbf{} &  &  &  &  &  &  &  &  &  &  &  &  &  &  &  &  &  &  &  & \checkmark \\ \hline 
Dynamic accumulator\newline \cite{133} &  &  &  & \textbf{} &  &  & \textbf{} & \textbf{} &  &  &  &  &  &  &  &  &  &  &  &  &  &  &  &  &  &  & \checkmark &  \\ \hline 
BBS+ signature\newline \cite{129} &  &  &  & \textbf{} &  &  & \textbf{} & \textbf{} &  &  &  &  &  &  &  &  &  &  &  &  &  &  &  &  &  & \checkmark &  &  \\ \hline 
Role-centric attribute-based access control \cite{46}  & \checkmark &  &  &  &  &  &  &  &  &  &  &  &  &  &  &  &  &  &  &  &  &  &  &  &  &  &  &  \\ \hline 
Private Stream Aggregation \cite{52} &  & \checkmark & \checkmark &  &  &  &  &  &  &  &  &  &  &  &  &  &  &  &  &  &  &  &  &  &  &  &  &  \\ \hline 
Homomorphic encryption \cite{54} &  &  &  & \checkmark &  &  &  &  &  &  &  &  &  &  &  &  &  &  &  &  &  & \checkmark &  &  & \checkmark &  &  &  \\ \hline 
Paillier encryption\newline   \cite{57} &  &  &  &  & \checkmark &  &  &  & \checkmark &  &  &  &  &  &  &  &  &  &  &  &  & \checkmark &  &  &  &  &  &  \\ \hline 
Boneh-Goh-Nissim cryptosystem \cite{60} &  &  &  &  &  & \checkmark & \checkmark &  &  &  &  &  &  &  &  &  &  &  &  &  &  &  &  &  &  &  &  &  \\ \hline 
Public key encryption with keyword search \cite{73} &  &  &  &  &  &  &  &  &  & \checkmark &  &  &  &  &  &  &  &  &  &  &  &  &  &  &  &  &  &  \\ \hline 
Hidden Vector Encryption \cite{75} &  &  &  &  &  &  &  &  &  &  & \checkmark &  &  &  &  &  &  &  &  &  &  &  &  &  &  &  &  &  \\ \hline 
Batch verification algorithm \cite{80} &  &  &  &  &  &  &  &  &  &  &  & \checkmark &  &  &  &  &  &  &  &  &  &  &  &  &  &  &  &  \\ \hline 
Identity-Committable Signature \cite{84} &  &  &  &  &  &  &  &  &  &  &  &  &  & \checkmark &  &  &  &  &  &  &  &  &  &  &  &  &  &  \\ \hline 
Partially blind signature\newline \cite{85} &  &  &  &  &  &  &  &  &  &  &  &  &  & \checkmark &  &  &  &  &  &  &  &  &  &  &  &  &  &  \\ \hline 
El-Gamal public key encryption \cite{99} &  &  &  &  &  &  &  &  &  &  &  &  &  &  &  & \checkmark &  &  &  &  &  &  &  &  &  &  &  &  \\ \hline 
Schnorr signature scheme \cite{100} &  &  &  &  &  &  &  &  &  &  &  &  &  &  &  & \checkmark &  &  &  & \checkmark & \checkmark &  &  &  &  &  &  &  \\ \hline 
Anonymous signature scheme \cite{101} &  &  &  &  &  &  &  &  &  &  &  &  &  &  &  &  & \checkmark &  &  &  &  &  &  &  &  &  &  &  \\ \hline 
Id-based signature scheme \cite{102} &  &  &  &  &  &  &  &  &  &  &  &  &  &  &  &  & \checkmark &  &  &  &  &  & \checkmark & \checkmark &  &  &  &  \\ \hline 
Off-line/online signature \cite{104} &  &  &  &  &  &  &  &  &  &  &  &  &  &  &  &  &  & \checkmark &  &  &  &  &  &  &  &  &  &  \\ \hline 
Elliptic curve cryptography \cite{107} &  &  &  &  &  &  &  &  &  &  &  &  &  &  &  &  &  &  & \checkmark & \checkmark &  &  &  &  &  &  &  &  \\ \hline 
Identity-based\newline Encryption \cite{115} &  &  &  &  &  &  &  &  &  &  &  &  &  &  &  &  &  &  &  &  &  &  &  & \checkmark &  &  &  &  \\ \hline 
Pedersen commitment\newline \cite{127} &  &  &  &  &  &  &  &  &  &  &  &  &  &  &  &  &  &  &  &  &  &  &  &  &  & \checkmark &  &  \\ \hline 
\end{tabular}}
\label{tab:Tab8}
\end{table}
\begin{table}[h]
         \centering
         \caption{The schemes that use Homomorphic encryption and Paillier encryption}
        \begin{tabular}{||p{0.5in}|p{1.1in}|p{2in}||} \hline 
       \textbf{Scheme} & \textbf{Type} & \textbf{Design goal} \\ \hline 
       Li et al. (2015) \cite{53} & RLWE-based somewhat homomorphic encryption \cite{54} & Proposing an optimized privacy-preserving dual-functional aggregation scheme \\ \hline 
       Chen et al. (2015) \cite{55} & Homomorphic Paillier cryptosystem \cite{57} & Proposing a privacy-preserving data aggregation scheme with fault tolerance \\ \hline 
       Li et al. (2014) \cite{69} & Homomorphic Paillier cryptosystem \cite{57} & Proposing a multi-keyword range query scheme \\ \hline 
       Dimitriou et al. (2016) \cite{94} & Homomorphic Paillier cryptosystem \cite{57} & Proposing two decentralized protocols to securely aggregate the measurements of $n$ smart meters \\ \hline 
       Han et al. (2016) \cite{118} & Partial homomorphic encryption \cite{176} & Proposing an integrated privacy-preserving data management architecture \\ \hline 
       \end{tabular}
        \label{tab:Tab8a}
        \end{table}
 \begin{figure}[h]
    \centering
    \includegraphics[width=0.5\linewidth]{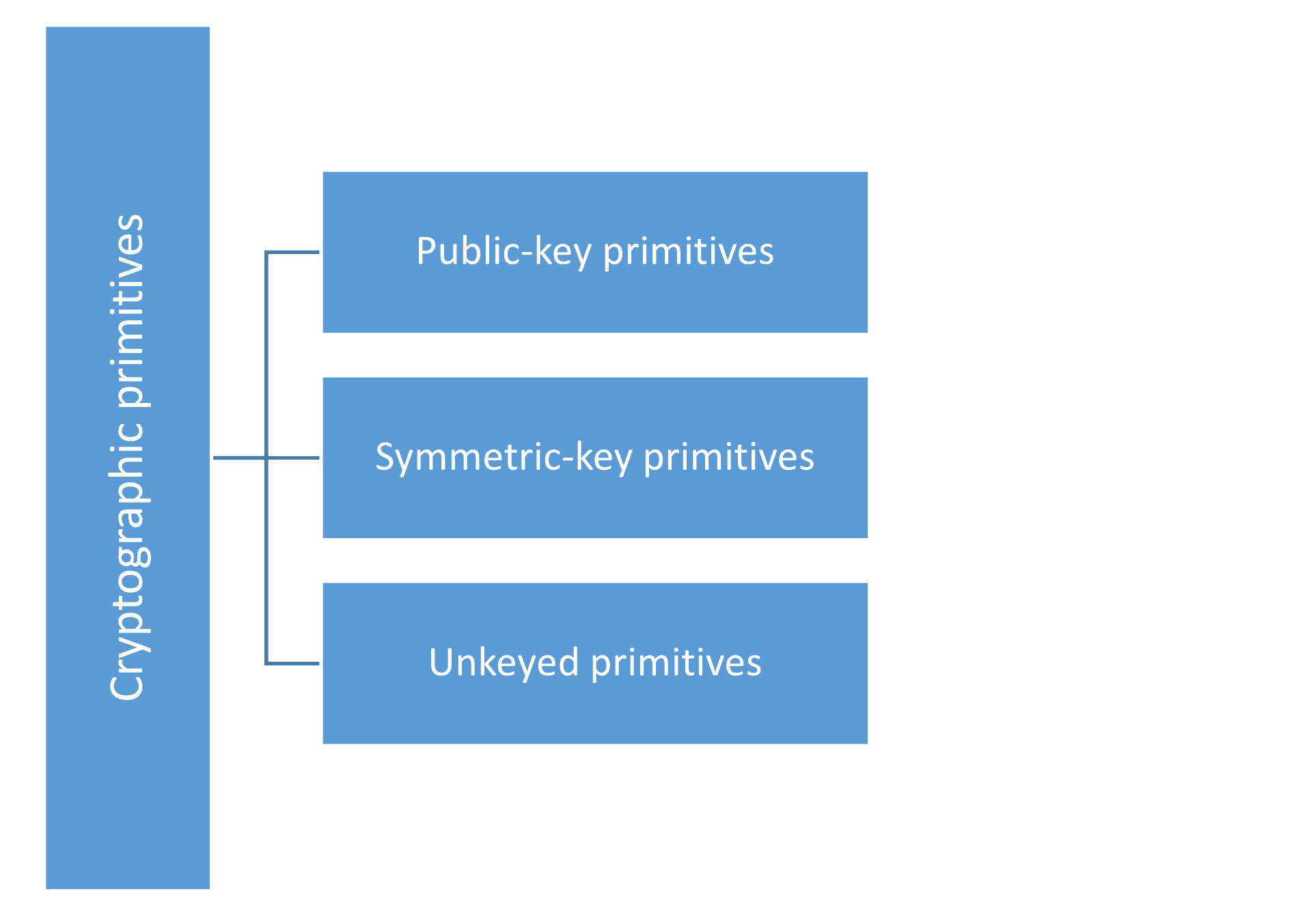}
    \caption{Taxonomy of cryptographic primitives}
    \label{fig:Fig4}
    \end{figure}
    \begin{table}[h]
     \centering
     \caption{Game theoretic and formal proof approaches used in privacy-preserving schemes for Smart Grids}
    \begin{tabular}{||p{0.5in}|p{1.5in}|p{2in}||} \hline 
    \textbf{Scheme} & \textbf{Approach} & \textbf{Main results} \\ \hline 
    Saxena et al. (2015) \cite{33} & BAN-Logic \cite{47} & Justification of security analysis \\ \hline 
    Jia et al. (2014) \cite{51} & Game on leaking individual meter's measurements & Define the formal attack model and privacy \\ \hline 
    Fan et al. (2014) \cite{76} & Sequences of games & Proof of semantic security;\newline Proof of unforgeability;\newline Proof of batch verification security \\ \hline 
    Gong et al. (2016) \cite{83} & Zero-Knowledge Proof \cite{86} & Ensures that adversaries cannot change the pseudonym in the message \\ \hline 
    Rahman et al. (2016) \cite{88} & Zero-Knowledge Proof \cite{86} & The winner can claim the incentive \\ \hline 
    Wan et al. (2016) \cite{89} & Zero-Knowledge Proof \cite{86} & Ensures anonymous authentication and rewarding;\newline Ensures unlinkable credentials and rewards \\ \hline 
    He et al. (2016) \cite{92} & Game played between a probabilistic polynomial time adversary \textit{A} and a challenger \textit{C} & Provide the mutual authentication and is semantically secure in the random oracle model \\ \hline 
    Dimitriou et al. (2016) \cite{94} & Measurement indistinguishability experiment & Proof the privacy of meter \textit{i}'s measurements \\ \hline 
    He et al. (2016) \cite{96} & Unforgeability, which the semantic security is defined through games played between a simulator and an adversary & Define the security model \\ \hline 
    Tsai et al. (2016) \cite{108} & Security proofs used in ref. \cite{116} & Proof the security model of ID-based multiservice provider authentication scheme \\ \hline 
    Liu et al. (2016) \cite{124} & Zero-Knowledge Proof  used in ref. \cite{128} & Proof the cryptographic building blocks \\ \hline 
    Saxena et al. (2016) \cite{125} & Strand space model \cite{132}  & Define the protocol as a sequence of events for each role of the electric vehicle, local aggregator, and certification/registration authority \\ \hline 
    \end{tabular}
    \label{tab:Tab9}
    \end{table}
    
\section{Countermeasures, game theoretic and formal proof approaches}\label{sec:countermeasures,-game-theoretic-and-formal-proof-approaches}
In order to satisfy the security requirements for secure Smart Grid communications, namely, authentication, integrity, non-repudiation, access control, and privacy, privacy-preserving schemes use the cryptography as a countermeasure. In this section, we will discuss these cryptographic methods and we provide a description for game theoretic and formal proof approaches used in privacy-preserving schemes for Smart Grids.

\subsection{Countermeasures}

In the literature, the cryptographic primitives are classified into three categories, namely, public-key primitives, symmetric-key primitives, and Unkeyed primitives, as shown in Fig. \ref{fig:Fig4} \cite{81}.  Note that we have not found any privacy-preserving scheme for Smart Grid that use symmetric-key primitives. In addition, most of the privacy-preserving schemes for Smart Grid use secure cryptographic hash function \cite{157}.

The cryptographic methods used in privacy-preserving schemes for Smart Grid are summarized in Tab. \ref{tab:Tab8}. For ensuring  confidentiality, authenticity and non-repudiability of Smart Grid communications, the privacy preserving schemes use a public key cryptography. The scheme in \cite{126} uses PASSERINE public key cryptosystem \cite{134}, which is a lightweight public key encryption mechanism. To encrypt a message $m$ using PASSERINE, a user squares it modulo the public modulus $n$: $Z\ =\ m^2\ ({\rm mod}\ n)$ . The scheme in \cite{125} uses a Dynamic accumulator \cite{133} which the accumulation function computes the accumulation value of prime numbers and the witness-generation function computes the witness in order to uses them in the authentication function. The scheme \cite{124} uses BBS+ signature \cite{129} which is based on the short group signature. BBS+ uses five algorithms, namely, \textit{GMSetup}, \textit{APSetup}, \textit{Join}, \textit{GrantingAccess}, and \textit{RevokingAccess}. The scheme \cite{118} uses role-centric attribute-based access control \cite{46} which the roles are assigned to users. The scheme \cite{118} uses role-centric attribute-based access control \cite{46} on user attributes, object attributes, and permission filtering policy. Both schemes \cite{35} and \cite{49} use private stream aggregation such as privacy-preserving aggregation of time-series data \cite{52}. Therefore, homomorphic encryption \cite{54} and paillier encryption \cite{57} are used by five schemes in particular \cite{53,55,69,94,118} as presented in Tab \ref{tab:Tab8a}. The homomorphic property under a message $x$ is as follows, $\left(x_1\right)\cdot \varphi \left(x_2\right)=x^e_1x^e_2\ {\rm mod}\ m={\left(x_1x_2\right)}^e{\rm mod\ }m={\rm \ }\varphi (x_1\cdot x_2)$ . In the Paillier encryption, the public key is the modulus $m$ and the base $g$, the encryption of a message $x$ is $\varphi \left(x\right)=g^xr^m\ {\rm mod\ }\ m^2$, for some random $r\in \{0,\cdots ,\ m-1\}$. The homomorphic property of Paillier encryption is as follows: $\varphi \left(x_1\right)\cdot \varphi \left(x_2\right)=\left(g^{x_1}{r_1}^m\right)\left(g^{x_2}{r_2}^m\right){\rm \ mod\ }m^2=g^{x_1+x_2}{\left(r_1r_2\right)}^m\ {\rm mod\ }m^2=\varphi (x_1+x_2)$. Both schemes \cite{58} and \cite{62} use Boneh-Goh-Nissim cryptosystem \cite{60} that is a homomorphic public-key encryption scheme. The scheme \cite{70} uses public key encryption with keyword search (PEKS) \cite{73}. The PEKS is based on four algorithms: 1) \textit{Setup} outputs public key and secret key, 2) \textit{PEKS} outputs a ciphertext \textit{C}, 3) \textit{Trapdoor} outputs the trapdoor information and 4) \textit{Test} for checking the ciphertext \textit{C}. The scheme \cite{71} uses hidden vector encryption (HVE) \cite{75} which can be viewed as an extreme generalization of anonymous identity-based encryption. The scheme \cite{76} uses batch verification algorithm \cite{80} that is based on a digital signature scheme. The scheme \cite{83} uses identity-committable signature \cite{84} and partially blind signature \cite{85} in order to secure against adaptive attacks. The scheme \cite{88} uses El-Gamal public key encryption \cite{99} and Schnorr signature scheme \cite{100}. Note that both schemes \cite{92,93} use Schnorr signature scheme \cite{100}. The scheme \cite{89} uses anonymous signature scheme \cite{101} which allows a signer's identity to be revealed if the signer signs the same message. The Id-based signature scheme \cite{102} is used by three schemes in particular \cite{89,96,108}. Off-line/online signature \cite{104} is used by the scheme \cite{90}. Using the algebraic structure of elliptic curves over finite fields, both schemes \cite{91,92} uses elliptic curve cryptography \cite{107}. Identity-based encryption \cite{115,170} is used by the scheme \cite{108}.

\subsection{Game theoretic and formal proof approaches}

In order to prove the effectiveness of their novel secure methods, researchers use game theoretic and formal proof approaches. Tab. \ref{tab:Tab9} presents the game theoretic and formal proof approaches used in privacy-preserving schemes for Smart Grids. In particular, Gong et al. \cite{83}, Rahman et al. \cite{88} and Wan et al. \cite{89} use Zero-Knowledge Proof \cite{86} as an approach to prove the schemes that can ensure anonymous authentication. In order to define the formal attack model and privacy, Jia et al. \cite{51} use a game on leaking individual meter's measurements. To prove the security model of ID-based multiservice provider authentication scheme, Tsai et al. \cite{108} uses security proofs of the work in \cite{116}. In order to prove a privacy-preserving scheme is secure in the random oracle model, He et al. in \cite{92} use a game played between a probabilistic polynomial time adversary \textit{A} and a challenger \textit{C}. Liu et al. \cite{124} uses Zero-Knowledge Proof presented in the work \cite{128} to prove the cryptographic building blocks. Based on strand space model \cite{132}, Saxena et al. \cite{125} defines the protocol as a sequence of events for each role of the electric vehicle, local aggregator, and certification/registration authority. Therefore, the idea of sequences of games \cite{158} is used by the scheme \cite{76} in order to prove the semantic security, unforgeability, and batch verification security.
 \begin{figure}[h]
    \centering
    \includegraphics[width=1\linewidth]{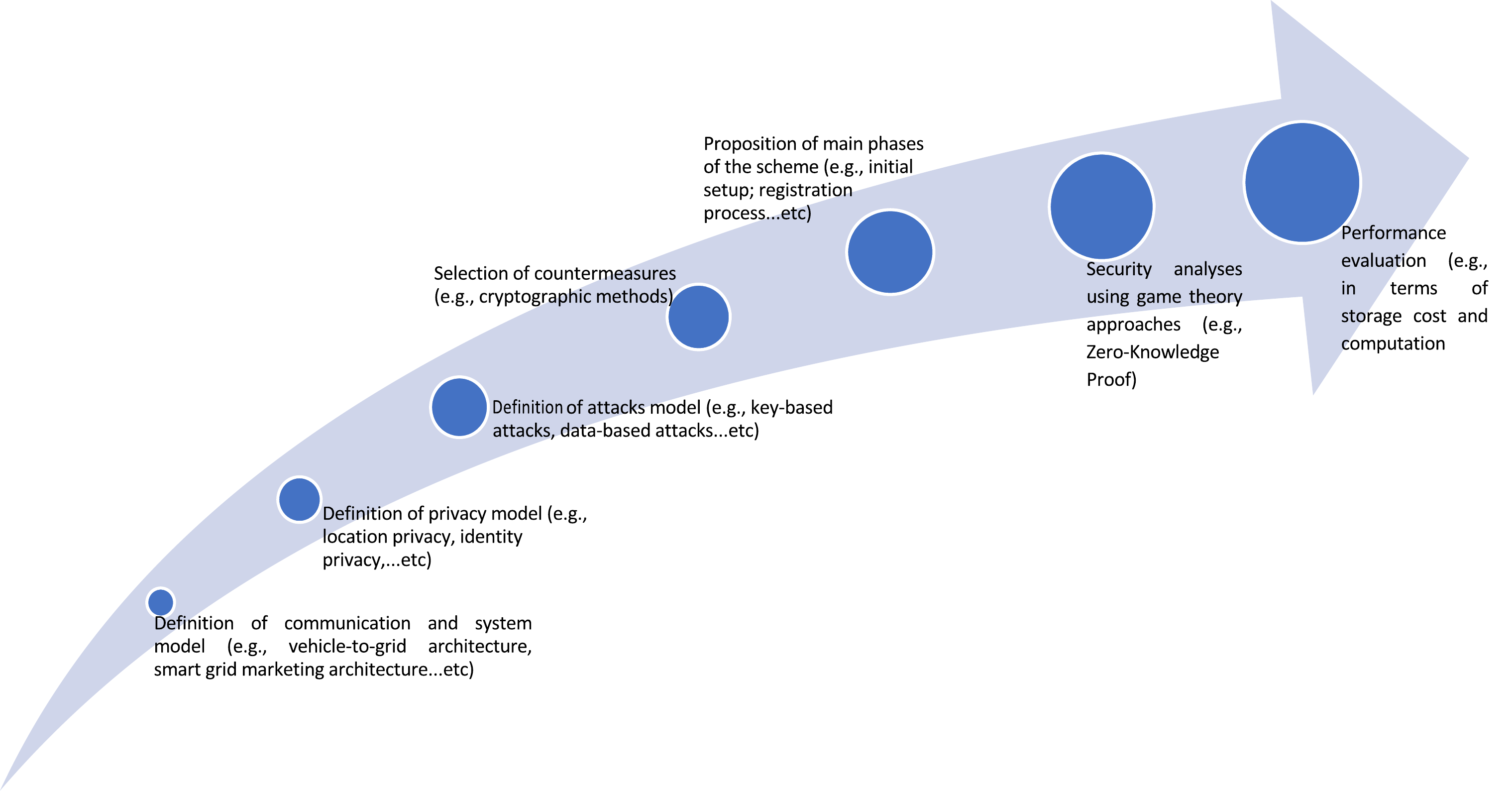}
    \caption{The realization processes of a privacy-preserving scheme for Smart Grids}
    \label{fig:Fig5a}
    \end{figure}
 \begin{figure}[h]
    \centering
    \includegraphics[width=0.6\linewidth]{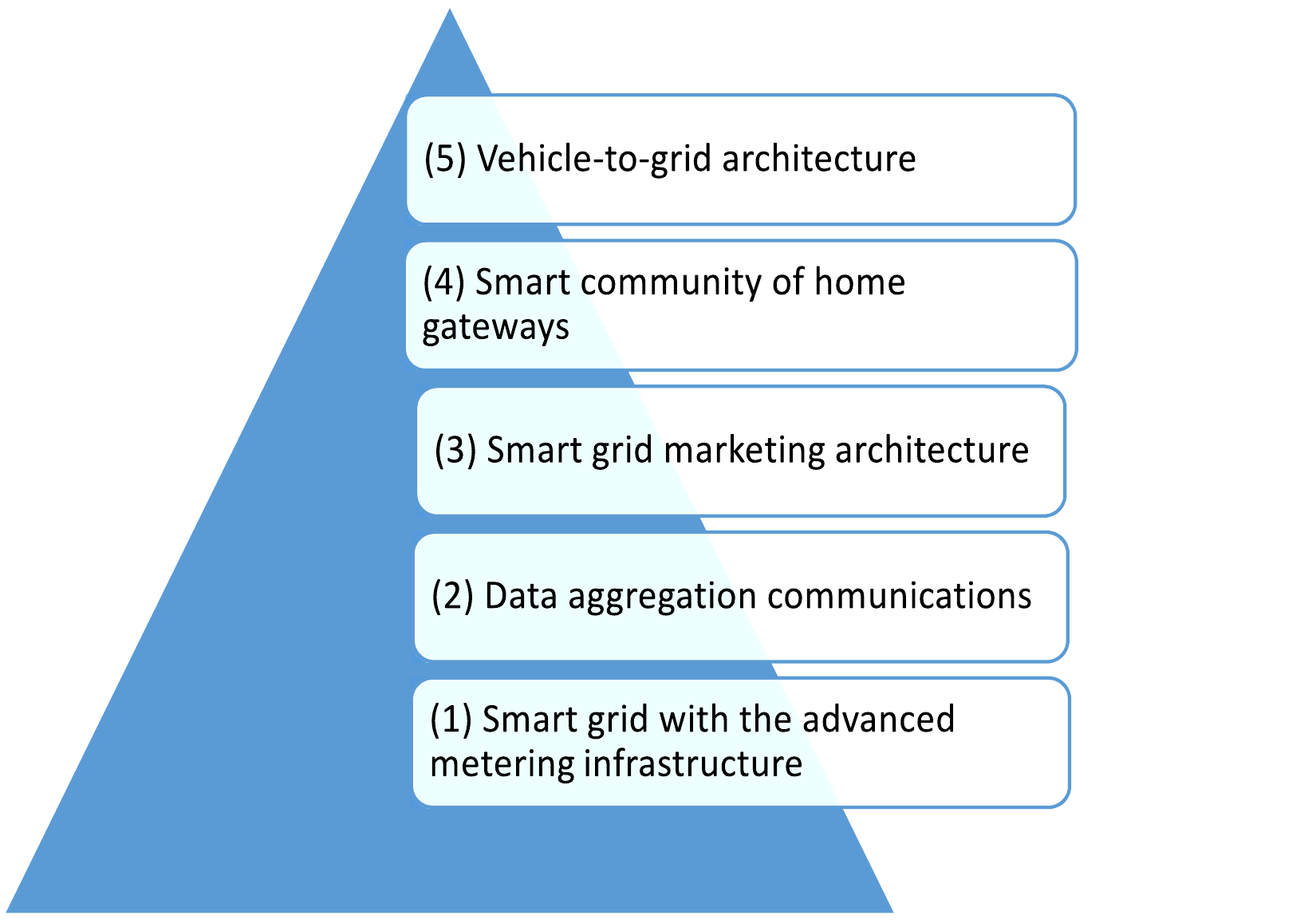}
    \caption{Communication and system models used for Smart Grids}
    \label{fig:Fig5}
    \end{figure}
\section{Privacy-preserving schemes for Smart Grids}\label{sec:privacy-preserving-schemes-for-smart-grids}
In this section, we in-detail examine thirty privacy-preserving schemes developed for or applied in the context of Smart Grids. As shown in Fig. \ref{fig:Fig5a}, the realization processes of a privacy-preserving scheme for Smart Grids are based on seven processes, namely, 1) definition of communication and system model (e.g., vehicle-to-grid architecture, smart grid marketing architecture...etc), 2) definition of privacy model (e.g., location privacy, identity privacy,...etc), 3) definition of attacks model (e.g., key-based attacks, data-based attacks...etc), 4) selection of countermeasures (e.g., cryptographic methods), 5) proposition of main phases of the scheme (e.g., initial setup; registration process...etc), 6) security analyses using game theory approaches (e.g., Zero-Knowledge Proof), and 7) performance evaluation (e.g., in terms of storage cost and computation complexity). In addition, as shown in Fig. \ref{fig:Fig5}, we classify these schemes in five categories based on the communication and system models, including, 1) Smart grid with the advanced metering infrastructure, 2) Data aggregation communications, 3) Smart grid marketing architecture, 4) Smart community of home gateways, and 5) Vehicle-to-grid architecture. The surveyed papers as shown in Tab. XIII are published between 2013 and 2016.

\begin{center}
\topcaption{Summary of privacy-preserving schemes for Smart Grids (Published between 2013 and 2016)}
\end{center}
{\small \begin{supertabular}{||p{0.5in}|p{0.9in}|p{0.9in}|p{0.9in}|p{1.2in}|p{1.5in}||} \hline 
\textbf{Scheme} & \textbf{Communication and system model} & \textbf{Privacy model} & \textbf{Goals } & \textbf{Main phases} & \textbf{Performances (+) and limitations (-)} \\ \hline 
\multicolumn{6}{|p{7in}|}{ \begin{center}
\textbf{ (1) Smart grid with the advanced metering infrastructure}
\end{center}} \\ \hline 
Gong et al. (2016) \cite{83} & Smart grid with the advanced metering infrastructure (AMI)\textbf{} & Customer privacy & Guarantees privacy, integrity, and availability & Registration process;\newline Metering and querying processes;\newline Settlement process;\newline Revocation process\textbf{} & + Efficient in anonymity\newline - No threat model presented\newline - No comparison with other schemes\newline - Needs study the communication and computation overheads\textbf{} \\ \hline 
Saxena et al. (2015)\newline \cite{33} & Smart grid with the advanced metering infrastructure (AMI) & Privacy of users \newline (employees, vendor engineers, maintenance personnel, and security officer) & Providing both authentication and authorization & Initial setup;\newline Identity creation;\newline Accessing device;\newline Verification of the Identities & + Efficient in terms of  communication and computation overheads compared to \cite{42,43}\newline + Can be applied to different user-roles\newline + Prevent various insider and outsider attacks\newline - Error-detection and fault tolerance are not considered \\ \hline 
Tan et al. (2016) \cite{87} & Smart grid with the advanced metering infrastructure (AMI) & Privacy of users\newline  & Guarantees privacy, integrity, and authenticity & Initialization;\newline Pseudo ID generation;\newline Message authentication & + Degree of anonymity vs. number of smart meters\newline + Computation overhead vs. number of smart meters\newline - No comparison with other schemes\newline - Considers only the false data attack \\ \hline 
Rahman et al. (2016) \cite{88} & Three entities, including, (a) energy supplier as registration manager, (b) automation server as bidding manager, and (c) bidders & Customer privacy & Providing anonymity, untraceability, non-linkability, no-impersonation, unforgeability, non-repudiation, verifiability, and integrity & Pre-processing stage;\newline Bidder registration;\newline Bidding setup \& key generation;\newline Bid verification;\newline Winner announcement;\newline Incentive claim & + Efficient in terms of computation and communication cost\newline - No comparison with other schemes\newline  \\ \hline 
\multicolumn{6}{|p{7in}|}{ \begin{center}
\textbf{ (2) Data aggregation communications}
\end{center}} \\ \hline 
Shi et al. (2015) \cite{35} & Data aggregation communications from the residential users to the control center in smart grid with a key management center (KMC) & Privacy of users \newline (residential users) & Providing privacy preservation, error-detection, and fault tolerance & System initialization;\newline Data encryption and reporting;\newline Aggregation with error detection;\newline Dynamic Join and Leave\newline  & + Accuracy of aggregation\newline + Efficient in communication and computation overheads compared to \cite{49,50}\newline - Many assumptions needed to understand implementation\newline - Considers only the malicious data mining attack \\ \hline 
Deng \newline et al. (2015)\newline \cite{36} & Smart distribution grid with a load-serving entity and multiple users (e.g., smart building/community or microgrid) & Privacy of users \newline  & Solving the demand response problem in a distributed manner & Subgradient projection;\newline Distributed algorithm; & + Comparison between dual decomposition and fast approach\newline - No threat model presented\newline - Many assumptions needed to understand implementation \\ \hline 
Sun et al. (2013)\newline \cite{49} & Data aggregation communications from the residential users to the control center in smart grid with a key management center (KMC) & Privacy of users \newline (residential users) & Providing privacy preservation, error-detection, and fault tolerance & System initialization;\newline Data encryption and reporting;\newline Aggregation with error detection & + Efficient in computation complexity and communication overhead\newline - No threat model presented\newline - No comparison with other schemes \\ \hline 
Jia et al. (2014)\newline \cite{51} & One aggregator and users (customers) equipped with an electricity smart meter & Privacy of data aggregation & Providing the data aggregation without leaking individual meters measurements & System initialization;\newline Meter's reading report;\newline Privacy-preserving aggregation & + Efficient in term of computation cost of the aggregator\newline + Consider the effect of the human factor on the data aggregate\newline - No comparison with other schemes \\ \hline 
Fan et al. (2014) \cite{76} & Smart grid with home area network (HAN), building area network (BAN), and industry area network (IAN);\newline Three entities, including, an aggregator, users, and an off-line trusted third party & Privacy of data aggregation & Avoiding the internal attacks & Initialization;\newline Registration;\newline Aggregation; & + Efficient in terms of aggregation and batch verification\newline + Ensuring data integrity compared to the schemes \cite{78,79,96}\newline - Energy cost is not considered\newline - Identity privacy and location privacy are not considered  \\ \hline 
Li et al.\newline (2015)\newline \cite{53}\newline \newline  & Three communication parties in smart grid, namely, data \& control center, residential gateway, and a large number of residential users & Privacy of users;\newline Privacy of data aggregation\newline  & Supporting dual-functional (mean and variance) aggregation under providing the data aggregation & System initialization;\newline User report generation;\newline Privacy-preserving report aggregation;\newline Secure report reading & + Efficient in term of computation cost and communication overhead\newline - No comparison with other schemes\newline - Considers only chosen-plaintext-attack \\ \hline 
Chen et al.\newline (2015)\newline \cite{55}\newline  & Smart grid communication architecture, including a trusted authority, a set of servers, a control center, a residential gateway, and a large number of residential users & Privacy of users;\newline Privacy of servers\newline Privacy of data aggregation\newline  & Providing the data aggregation with fault tolerance & System initialization;\newline User report generation;\newline Privacy-preserving report aggregation;\newline Secure report reading;\newline Fault tolerance handling & + Efficient in term of communication overhead\newline - Computation cost is not studied\newline - Compared only with the scheme \cite{56} \\ \hline 
Chen et al.\newline (2015)\newline \cite{58} & Smart grid communication architecture, including a trusted authority, a set of servers, a control center, a residential gateway, and a large number of residential users & Differential privacy of users;\newline Privacy of data aggregation\newline  & Achieving privacy-preserving aggregation of multiple functions such as average, variance...etc & System initialization;\newline User report generation;\newline Privacy-preserving report aggregation;\newline Secure report reading & + Efficient in terms of computation complexity and communication overhead\newline - Compared only with the scheme \cite{52}\newline - Many assumptions needed to understand implementation \\ \hline 
Bao et al.\newline (2015)\newline \cite{62} & Smart grid communication architecture, including a trusted authority, a set of servers, a control center, a residential gateway, and a great number of residential users & Differential privacy of data aggregation\newline  & Ensuring differential privacy of data aggregation with fault tolerance & System initialization;\newline Data aggregation request;\newline Data aggregation request relay;\newline User report generation;\newline Privacy-preserving report aggregation;\newline Secure report reading & + Efficient in terms of storage cost and computation complexity\newline + Utility of differential privacy\newline + Robustness of fault tolerance\newline - Communication overhead is not studied\newline - Data integrity is not considered \\ \hline 
Wang et al. (2016) \cite{91} & Five different entities, including, users, public cloud server, smart meters, utility provider, and trusted third party & Identity privacy & Balances the anonymity and traceability for outsourcing small-scale data linear aggregation & Setup; Enc; TTP-Dec; Re-key; LiAgg-ReEnc; UPDec & + Efficient in terms of Linear aggregation and confidentiality compared to the RVK scheme \cite{105} and the LMO scheme \cite{106}\newline - Location privacy is not considered \\ \hline 
He et al. (2016) \cite{96} & Three participants: an offline third trusted party, an aggregator, and a user & Privacy of users;\newline Privacy of data aggregation\newline  & Achieving authentication  and privacy-preserving data aggregationagainst internal attackers & Initialization;\newline Registration;\newline Aggregation; & + Efficient in term of computational cost compared to the scheme \cite{76}\newline + Internalattacks are not considered\newline - Insider attacks are not considered\newline - Energy cost and locationprivacy are not considered \\ \hline 
\multicolumn{6}{|p{7in}|}{ \begin{center}
\textbf{ (3) Smart grid marketing architecture}
\end{center}} \\ \hline 
Jiang et al. (2015) \cite{64} & SCADA in smart grid with three main components, namely, human--machine interface, master terminal unit, and remote terminal unit & Availability under privacy of users;\newline Forward secrecy;\newline Backward secrecy;\newline  & Solving the availability problem in resource-constrained SCADA system & System initialization;\newline Rekeying;\newline Self-healing mechanism;\newline Adding new member nodes;\newline Re-initialization mechanism & + Minimize the computation, memory, communication,\newline and energy costs\newline + Efficient compared to the schemes \cite{65,66}\newline - Many assumptions about the privacy needed to understand implementation \\ \hline 
Li et al.\newline (2014)\newline \cite{69} & Smart grid marketing architecture with main four parts, including, electricity generators, retailers, data center, and filtering center & Query privacy\newline  & Providing secure transactions between sellers and buyers;\newline Achieving confidentiality of keywords, authentication, data integrity and query privacy & System initialization;\newline Auction message creating;\newline Trapdoor aggregating;\newline Filtering & + Can achieving ranked search and personalized search simultaneously compared to \cite{70} and \cite{71}\newline + Efficient in terms of computation and communication overhead\newline - No threat model presented\newline - Energy costs is not considered \\ \hline 
Wen et al . (2014) \newline \cite{70}\newline  & Smart grid marketing architecture with main three parts, including, energy sellers, energy buyers, and auction managers (with two servers: a registration server (RS) and an auction server) & Privacy of the energy buyers & Achieving privacy of the energy buyers, bid integrity, and prefiltering ability & Registration;\newline Bidding;\newline Pre-filtering;\newline Decision-of-winner & + Efficient compared to the scheme \cite{72} in terms of the computation and communication overhead in the one keyword search process\newline - No threat model presented\newline - Ranked search and personalized search are not considered \\ \hline 
Wen et al . (2013)\newline \cite{71}\newline  & Residential area composed of a  control center, two cloud servers, a requester and some residential users & Query privacy\newline  & Providing the data confidentiality and privacy by introducing an HVE technique & Construction of the range query predicate;\newline Encrypted data deposit;\newline Range query & + Efficient compared to the scheme \cite{74} in terms of communication overhead, computation complexity and response time\newline - No threat model presented\newline  \\ \hline 
\multicolumn{6}{|p{7in}|}{ \begin{center}
\textbf{ (4) Smart community of home gateways}
\end{center}} \\ \hline 
Liang et al. (2013) \cite{77} & Homogeneous smart community consisting of home gateways & Identity privacy;\newline location privacy\newline  & Enables a resident to send a service request to nearby homes under the identity privacy and the location privacy & Proximity score calculation;\newline Communication phase of users;\newline Communication phase of homes & + Efficient in service rate and obtained bandwidth\newline - No threat model presented\newline - No comparison with other schemes \\ \hline 
Bao et al. (2016) \cite{90} & A trusted authority, a control center, a residential gateway, and a great number of residential users U & Privacy of users\newline  & Guarantees privacy and data integrity simultaneously & System initialization;\newline Session key agreement;\newline Autonomous cluster formation;\newline Off-line pre-computation;\newline Online electricity usage report;\newline Privacy-preserving report collection;\newline Secure report reading; & + Efficient in terms of computation cost and communication overhead compared to both the schemes \cite{103,76}\newline - Insider attacks are not considered \\ \hline 
He et al. (2016) \cite{92} & A smart meter, a server provider and a trusted third party & Customer privacy & Providing the smart meter anonymity and mutual authentication & System setup phase;\newline Extraction phase;\newline Key distribution phase; & + Efficient in terms of computation cost and communication cost compared to the scheme \cite{108}\newline + Resistance to impersonation attack, replay attack, modification attack, and man-in-the-middle attack\newline - Many assumptions needed to understand implementation \\ \hline 
Ni et al. (2016) \cite{93} & Three entities, including, control center,gateways, and smart meters\newline \newline  & Privacy of users\newline  & Guarantees the integrity of consumption reports & System setup;\newline User initialization;\newline Report generation;\newline Report aggregation;\newline Report reading; & + Efficient in terms of communication overhead and computational overhead compared to the schemes \cite{109,110,76,111}\newline - Energy costs is not considered \\ \hline 
Dimitriou et al. (2016) \cite{94} & A collection of smart meters with the utility provider/aggregator & Privacy of participating meters & Providing secure aggregate collected measurements & Initialization phase;\newline Non-interactive proof of plaintext equality;\newline Encryption \& Decryption; & + Efficient in terms of communication overhead and throughput\newline - No comparison with other protocols \\ \hline 
Tsai et al. (2016) \cite{108} & A set of smart meters and service providers in a smart grid & Anonymity of smart meter & Achieving mutual authentication and smart meter anonymity at the same time & System setup;\newline Smart meter extraction;\newline Service provider extraction;\newline Mutual authentication; & + Efficient in terms of computation costs compared to the schemes \cite{112,113,114}\newline -Data integrity and location privacy are not considered\newline  \\ \hline 
\multicolumn{6}{|p{7in}|}{ \begin{center}
\textbf{ (5)Vehicle-to-grid architecture}
\end{center}} \\ \hline 
Wan et al. (2016) \cite{89} & Vehicle-to-grid architecture\newline (consisting of electric vehicles, aggregators, and a trusted third party & Privacy of electric vehicles (EVs) owners & Providing the anonymous authentication and rewarding, unlinkable credentials and rewards, fair exchange for service rewarding, non-repudiation, efficient system maintenance & Initialization;\newline EV's registration by the trusted third party;\newline EV's anonymous authentication;\newline Local aggregators service rewarding to EV;\newline Maintenance for EVs; & + Efficient in processing time and online processing time of each phase\newline - No comparison with other schemes\newline - Identity privacy and location privacy are not considered \newline  \\ \hline 
Han et al. (2016) \cite{118}\newline  & Local aggregators in vehicle-to-grid network & Privacy of users\newline  & Protect the privacy of users during the whole data collection, aggregation, and publication processes & Data aggregation process;\newline Data publication process;\newline Peer-level distributed access control;\newline Separable key-chaining management;\newline Hierarchical concealed data aggregation; & + Efficient in aggregation throughput\newline + Efficient against replay attack, impersonation attack, chosen ciphertext attack, known plaintext attack, physical attack, and inference attack\newline - No comparison with other schemes \\ \hline 
Liu et al. (2016) \cite{124}\newline \newline  & Three parties, namely the user, the reservation service\newline provider, and the charging station & Privacy for users;\newline Location privacy & Providing the privacy-preserving reservation system for charging station & Setup phase;\newline User registration;\newline Reservation;\newline Charging; & + Efficient in communication overhead\newline - No comparison with other schemes\newline - Storage cost not considered\newline - IP hijacking, distributed denial-of-service attack and man-in-the-middle attack are not considered \\ \hline 
Saxena et al. (2016) \cite{125} & Three main entities, namely,  the electric vehicles, local aggregator, and certification/registration authority & Forward privacy;\newline Vehicle identity anonymity; \newline Vehicle untraceability and information privacy & Provides mutual authentications between the electric vehicles and the local aggregator, and between the electric vehicle and the certification/registration authority;\newline Preserves privacy of the electric vehicle's identity, location, charge/discharge selection, expected time, and battery status\newline  & Initial setup;\newline Electric vehicle registration;\newline Home-certification/registration authority and Home-local aggregator communication;\newline Scheme execution; & + Efficient in terms of computation and communication overhead compared to the schemes \cite{130,131}\newline + Vehicle identity anonymity and location privacy are considered\newline - Energy cost is not considered  \\ \hline 
Abdallah et al. (2016) \cite{126}\newline  & Six main entities, namely, electric vehicles, local aggregators, an access point, charging\newline stations, a trusted authority & Privacy of electric vehicles;\newline Location privacy; & Guarantee the authentication, accountability, confidentiality, and integrity & Initialization phase;\newline Operation phase;\newline Billing phase; & + Efficient in terms of communication overhead and computation complexity compared to the scheme \cite{131}\newline - Forward privacy is not considered  \\ \hline 
\end{supertabular}}

 \begin{figure}[h]
    \centering
    \includegraphics[width=0.7\linewidth]{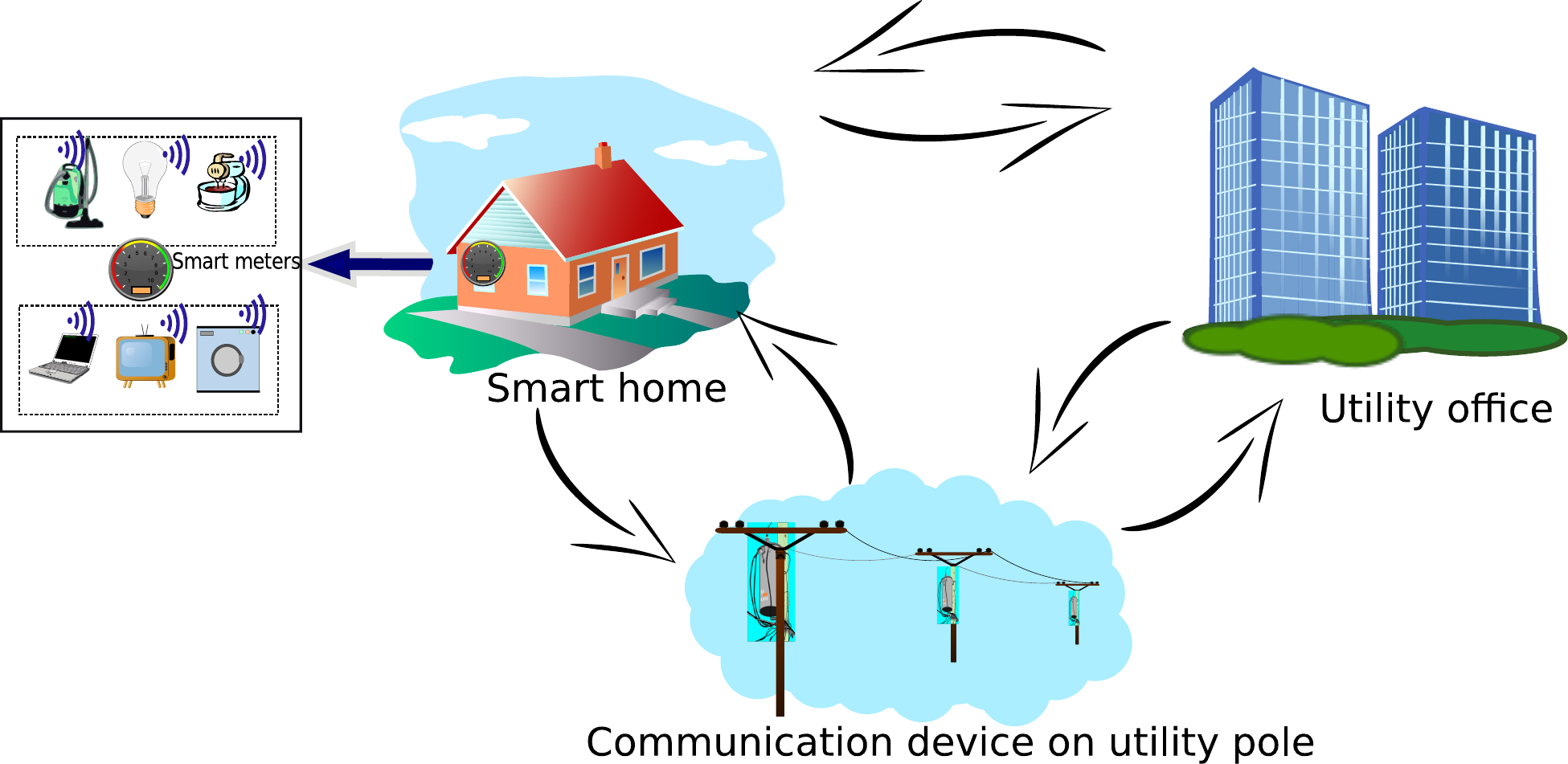}
    \caption{Advanced metering infrastructure}
    \label{fig:Fig6}
    \end{figure}
\subsection{Smart grid with the advanced metering infrastructure}
The advanced metering infrastructure (AMI) depends on sophisticated metering devices referred to as smart meters, as discussed in the work of Karnouskos et al. in \cite{136}. Fig. \ref{fig:Fig6}. shows an example of an AMI infrastructure which is based on three parties, namely, a smart home, a communication device on utility pole and a utility office. 
Saxena et al. \cite{33} proposes a scheme in order to provide the privacy of users (employees, vendor engineers, maintenance personnel, and security officers) in a Smart Grid with the advanced metering infrastructure (AMI). Based on three phases, namely, 1) identity creation, 2) accessing device, and 3) verification of the identities, the scheme that is proposed in \cite{33} is not only efficient in terms of communication and computation overhead, compared to the works in \cite{42,43}, but also can be applied to different user-roles and can prevent various insider and outsider attacks. Error-detection and fault tolerance are not considered in the scheme \cite{33} compared to the scheme \cite{35}. Deng et al. \cite{36}, proposes an idea to solve the demand response problem with both spatially- and temporally-coupled constraints in the smart distribution grid with a load-serving entity and multiple users. In addition, a recent work presented in \cite{83}, where Gong et al. proposes an idea in order to guarantee simultaneously privacy, integrity, and availability in smart grid with the advanced metering infrastructure. Based on three main processes, including, 1) metering and querying process; 2)settlement process; and 3) revocation process, the scheme can preserve the customer privacy by ensuring the anonymity of fine-grained metering data. Similarly to the scheme in \cite{83}, Tan et al. in \cite{87} proposes a pseudonym-based privacy-preserving scheme which is capable of detecting false data injection attacks in a smart-grid system which is equipped with advanced metering infrastructure (AMI). Besides, the scheme in \cite{87} can reassure privacy, integrity, and authenticity.

Rahman et al. \cite{88} proposes a scheme which considers three entities in a smart grid, including, (a) energy supplier as registration manager, (b) automation server as bidding manager, and (c) bidders. Specifically, the proposed scheme focuses on secure and private bidding for these three entities without relying on any trusted third party. Based on two main stages, namely, 1) winner announcement and 2) incentive claim, the scheme in \cite{88} can provide anonymity, untraceability, non-linkability, no-impersonation, unforgeability, non-repudiation, verifiability, and integrity. In addition, the scheme presented in \cite{88} is efficient in terms of computation and communication cost, but lacks comparison with other schemes.

\begin{table}[h]
\centering
\caption{Summary of the basic characteristics of the privacy-preserving schemes of data aggregation for Smart Grids}
{\footnotesize
\begin{tabular}{||p{0.6in}|p{0.45in}|p{0.45in}|p{0.45in}|p{0.45in}|p{0.45in}|p{0.45in}|p{0.45in}|p{0.45in}|p{0.45in}|p{0.45in}||} \hline 
\textbf{} & \multicolumn{10}{|p{4in}||}{\begin{center}
\textbf{Privacy-preserving schemes of data aggregation for Smart Grids}
\end{center}} \\ \hline 
\textbf{Char.} & \textbf{\cite{35}} & \textbf{\cite{49}} & \textbf{\cite{51}} & \textbf{\cite{76}} & \textbf{\cite{53}} & \textbf{\cite{55}} & \textbf{\cite{58}} & \textbf{\cite{62}} & \textbf{\cite{91}} & \textbf{\cite{96}} \\ \hline 
Aggregation method\newline  & Private stream aggregation\cite{52} & Private stream aggregation\newline \cite{52} & Cumulative information of usage patterns as the aggregators & Aggregation tree construction & Summation of total homomorphic multiplications & Aggregated data is computed and reported by a residential gateway & Average aggregation;\newline Variance aggregation;\newline One-way ANOVA aggregation & Aggregate the measurements into an integrated one & Linear aggregation & Aggregation with authentication \\ \hline 
Overhead to setup/maintain the aggregation structure \newline &  High &  High & Medium &  Medium &  High &  Medium &  High &  Medium & Medium & Medium \\ \hline 
Scalability\newline  & Medium & Medium & High & Medium & High & High & High & Medium & Medium & Medium \\ \hline 
Resilience in case of node mobility & Low & Medium & Medium & Low & Medium & Low & Low & Medium & Low & Low \\ \hline 
Timing strategy\newline  & Periodic  & Periodic & Periodic & Regular interval & Periodic & Every 15 minutes & Every 15 minutes & Every 15 minutes & Periodic & Periodic \\ \hline 
Resilience to link failures & Medium & Low & High & High & Medium & Medium & Medium & High & Medium & High \\ \hline 
Total communication in smart grid & \textit{O(N)} & \textit{O(N)} & N/A & N/A & $\frac{n\ {\rm log}\ q}{l}$ & $(2n+2)\ •S_N$ & $(n\ +\ 1)\ •\ L_N$ & N/A & N/A & 1568 bits \\ \hline 
\end{tabular}}
\label{tab:Tab10}
\end{table}
 \begin{figure}[h]
    \centering
    \includegraphics[width=0.7\linewidth]{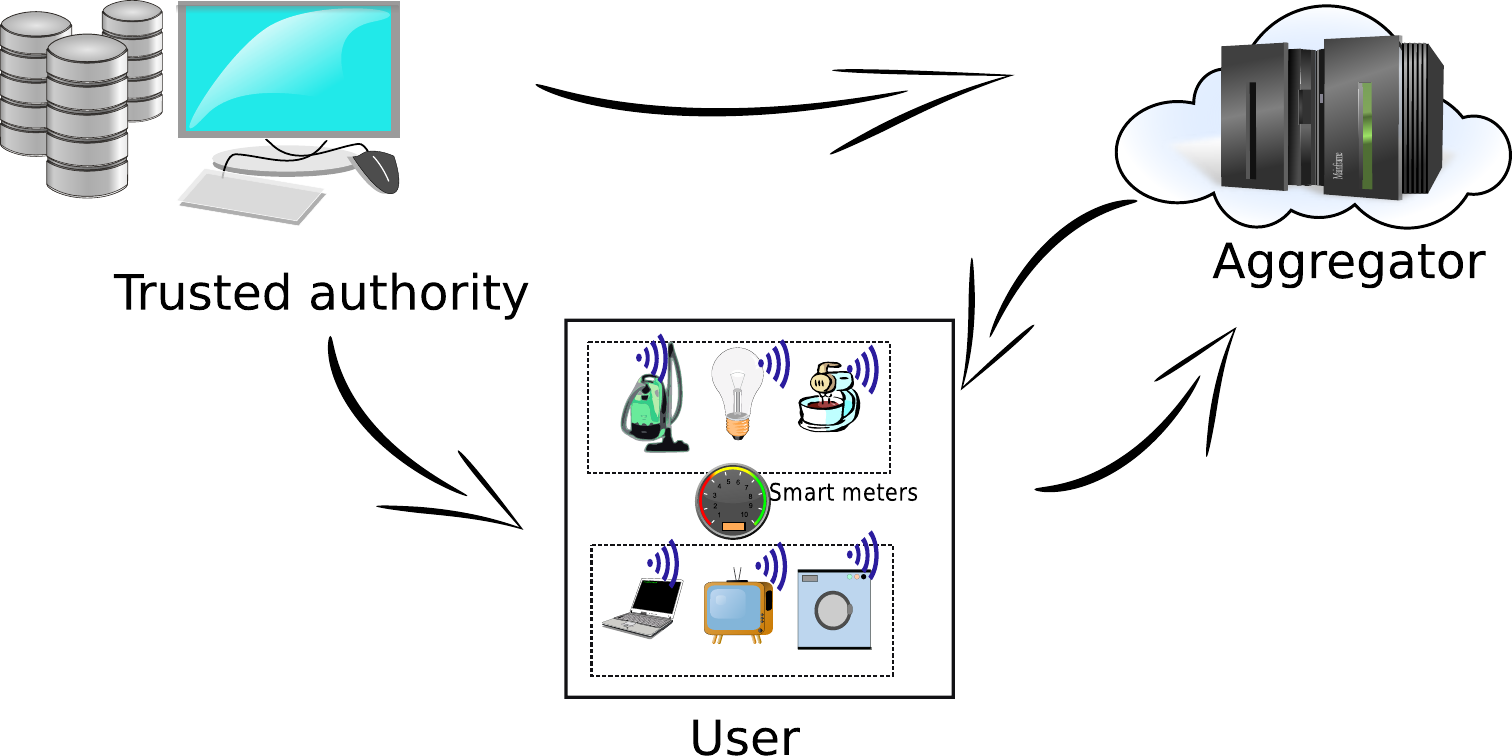}
    \caption{System model used in data aggregation communications}
    \label{fig:Fig7}
    \end{figure}
\subsection{Data aggregation communications}
Data aggregation techniques in wireless sensor networks have been proposed in many works \cite{137}. The basic idea of these techniques is based on using an aggregator connected with users and a trusted authority as presented in Fig. \ref{fig:Fig7}. However, privacy of data aggregation demonstrates many research challenges in privacy protection for smart grids, as discussed by Erkin et al. in \cite{61} and by Lu in \cite{63}. Tab. \ref{tab:Tab10} summarizes the basic characteristics of the privacy-preserving schemes of data aggregation for Smart Grids.

In order to preserve the privacy of residential users using data aggregation from the residential users to the control center in Smart Grid, Sun et al. \cite{49} proposes a protocol called APED. In order to improve the APED protocol, Shi et al. \cite{35} proposes a protocol, called DG-APED. Specifically, DG-APED protocol uses three main phases, namely, 1) data encryption and reporting, 2) aggregation with error detection and 3) dynamic Join and leave. Using both data encryption and reporting phase, each user perturbs his/her sensed data with generated noise. In addition, DG-APED is not only providing error-detection and fault tolerance, but also is efficient in terms of communication and computation overhead compared to the schemes in \cite{49,50}.

Similarly to the scheme in \cite{49}, Jia et al. \cite{51} considered one aggregator and users (customers) equipped with an electricity smart meter. Based on two phases, namely 1) meter's reading report and 2) privacy-preserving aggregation, the scheme in \cite{51} can provide data aggregation without leaking individual meters measurements. Another interesting work for privacy of data aggregation is presented in \cite{53}, where Li et al. proposes a scheme, called PDA. The PDA scheme is based on three phases, namely, 1) user report generation, 2) privacy-preserving report aggregation, and 3) secure report reading. PDA is efficient in terms of computation cost and communication overhead. In order to support both spatial and temporal aggregation of user electricity usages, Chen et al. \cite{55} proposes a scheme, called PDAFT, which is efficient in term of communication overhead compared to the scheme in \cite{56}, but lacks a study of computation cost. In the same context of PDAFT, Bao et al. \cite{62} proposes a scheme called DPAFT, which is a new differentially private data aggregation scheme with fault tolerance in order to provide fault tolerance for smart metering. Chen et al. in \cite{58} proposes a scheme called MuDA, which is similarly to the PDAFT scheme \cite{55} and the DPAFT scheme \cite{62}. The difference between PDAFT, DPAFT and MuDA is in cryptographic methods used, i.e., PDAFT uses the Paillier encryption \cite{57} and both DPAFT and MuDA use the Boneh-Goh-Nissim cryptosystem \cite{60}.

In order to avoid internal attacks against privacy of data aggregation, Fan et al. \cite{76} proposes an idea based on aggregation tree construction by running a breadth-first spanning tree algorithm. The scheme in \cite{76} ensures not only data integrity compared to the schemes \cite{78,79}, but is also efficient in terms of aggregation and batch verification. Similarly to the scheme in \cite{76}, He et al. in \cite{96} proposes a scheme that can achieve authentication and privacy-preserving data aggregation against internal attackers. Based on three main phases, namely, 1) initialization, 2) registration, and 3) aggregation, the scheme in \cite{96} is efficient in terms of computational cost compared to the scheme \cite{76}, but both energy cost and location privacy aspects are not considered. Wang et al. in \cite{91} proposes a protocol called BAT-LA in order to balance the anonymity and traceability for outsourcing small-scale data linear aggregation. The BAT-LA protocol is efficient in terms of linear aggregation and confidentiality compared to the RVK scheme \cite{105} and the LMO scheme \cite{106}.

 \begin{figure}[h]
    \centering
    \includegraphics[width=0.7\linewidth]{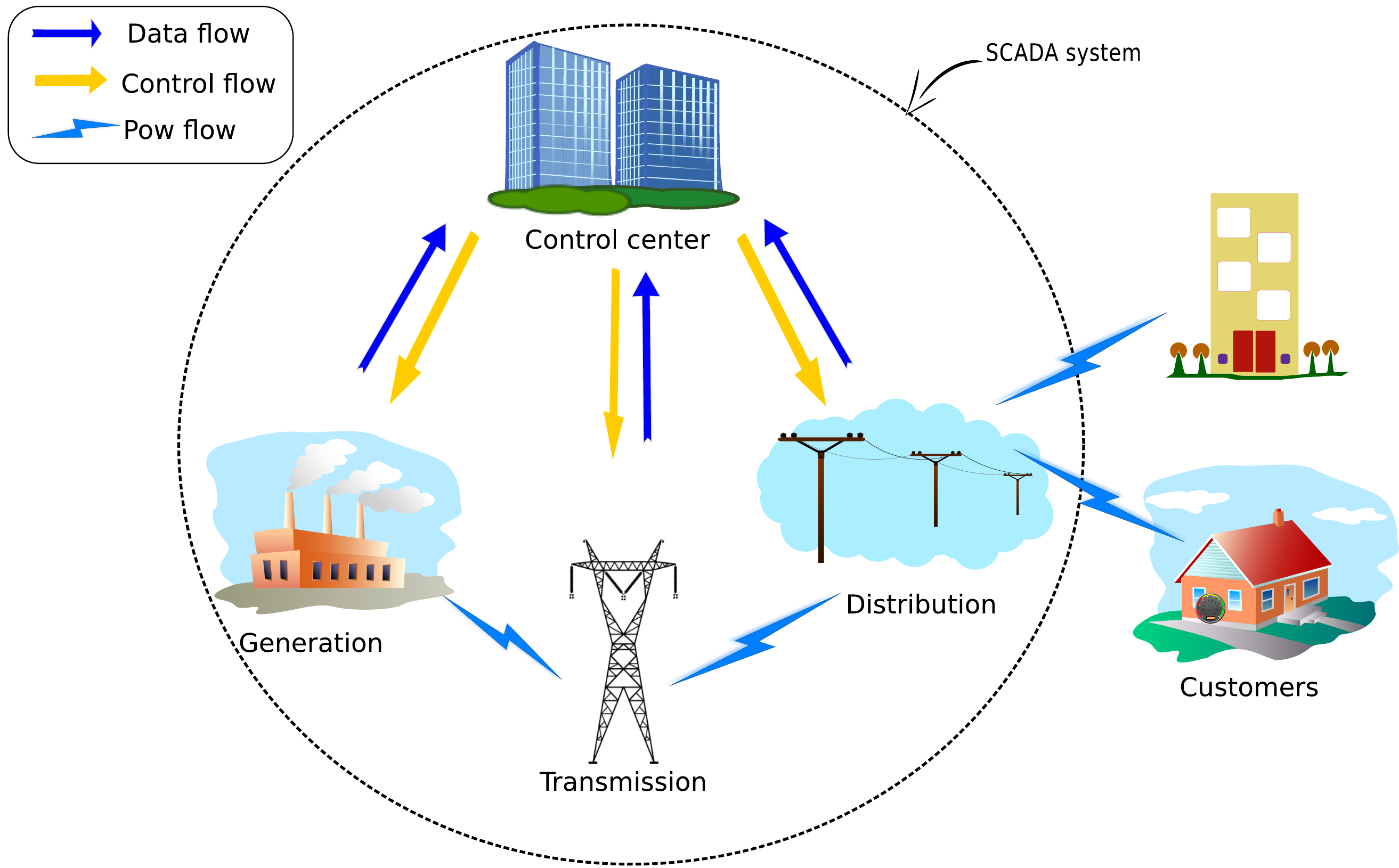}
    \caption{System model used in smart grid marketing architecture}
    \label{fig:Fig8}
    \end{figure}
\subsection{Smart grid marketing architecture}

The integration of Supervisory Control and Data Acquisition (SCADA) systems in Smart Grids allow the utility manager to remotely monitor and control networks (e.g. HANs (Home Area Networks), the BANs (Building Area Networks), the IANs (Industrial Area Networks), the NANs (Neighborhood Area Networks), the FANs (Field Area Networks), and the WANs (Wide Area Networks) \cite{1}. Malicious data in a SCADA system disrupt the management of these networks \cite{139,140}. For more details about the threats, risks and mitigation strategies in the area of SCADA security, we refer the reader to the survey \cite{142}. According to the conceptual model of National Institute for Standards and Technology (NIST), Fig. \ref{fig:Fig8} shows an example of a system model used in smart grid marketing architecture, which is based on four main components, namely, generation, transmission, distribution, and customer, feature two-way power and information flows in Smart Grid.

Jiang et al. in \cite{64} considered the availability under privacy of users in Smart Grid with three main components, namely, human--machine interface, master terminal unit, and remote terminal unit. Based on the dual directional hash chains, Jiang et al. proposes a scheme called LiSH+ to solving the availability problem in resource-constrained SCADA system. Specifically, LiSH+ uses four main phase, namely, 1) rekeying; 2) self-healing mechanism, 3) adding new member nodes, and 4) re-initialization mechanism. In addition, LiSH+ can minimize the computation, memory, communication, and energy costs compared to the schemes \cite{65,66}, but many assumptions about the privacy are needed in order to understand implementation.

Wen et al. \cite{70} proposes a scheme called SESA which considers  privacy of the energy buyers in Smart Grid marketing architecture with main three parts, including, energy sellers, energy buyers, and auction managers (with two servers: a registration server (RS) and an auction server). Based on three main phases, namely, bidding, pre-filtering, and decision-of-winner, the SESA scheme is efficient compared to the scheme in \cite{72} in terms of computation and communication overhead in the one keyword search process. Ranked search and personalized search are not considered in the SESA scheme. As a solution for query privacy, Wen et al.in  \cite{71} proposes a scheme called PaRQ for a typical residential area composed of a  control center, two cloud servers, a requester and some residential users. Based on three main phases, namely, construction of the range query predicate, encrypted data deposit, range query, PaRQ can provide the data confidentiality and privacy by introducing an HVE technique \cite{75}. In addition, PaRQ is efficient compared to the scheme \cite{74} in terms of communication overhead, computation complexity and response time, but no threat model is presented. Similarly to PaRQ scheme in the context of query privacy, Li et al. in \cite{69} proposes a scheme called EMRQ in Smart Grid marketing architecture with main four parts, including, electricity generators, retailers, data center, and filtering center. The EMRQ scheme can provide not only secure transactions between sellers and buyers, but can also achieve confidentiality of keywords, authentication, data integrity and query privacy. In addition, EMRQ can achieve ranked search and personalized search simultaneously compared to \cite{70} and \cite{71}.
 \begin{figure}[h]
    \centering
    \includegraphics[width=0.7\linewidth]{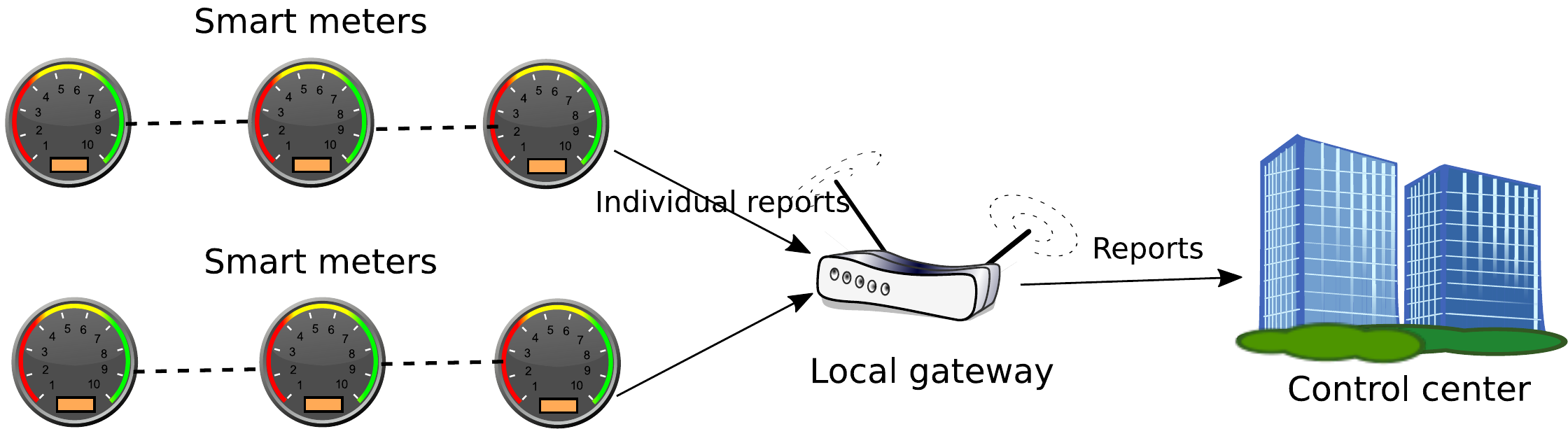}
    \caption{System model used in smart community of home gateways}
    \label{fig:Fig9}
    \end{figure}
\subsection{Smart community of home gateways}

A smart community of home gateways is a virtual environment composed of networked smart homes located in a local geographic region as discussed by Li et al. in \cite{138}. Fig. \ref{fig:Fig9} shows an example of a system model used in smart community of home gateways which is based on three parties, namely, smart meters, local gateway, and control center.

As discussed in the recent survey of Fadel et al. in \cite{82}, the Smart Grid network can be divided into three segments, including, Home Area Networks (HANs), Neighborhood Area Networks (NANs) and Wide Area Networks (WANs). Therefore, the identity privacy and the location privacy are the most important models for privacy in ad hoc networks, as discussed in the survey of Ferrag et al. in \cite{81}. Besides, Liang et al. in \cite{77} considered a homogeneous smart community consisting of home gateways, which can be considered as an ad hoc network. Specifically, Liang et al. in \cite{77} proposes a scheme called EPS in order to preserve both the identity privacy and the location privacy. Based on three main phases, namely, 1) proximity score calculation, 2) communication phase of users, and 3) communication phase of homes, the EPS scheme enables a resident to send a service request to nearby homes under the identity privacy and the location privacy. EPS is efficient in terms of service rate and obtained bandwidth, but needs further study to determine its feasibility in terms of computation and communication overhead. Bao et al. in \cite{90} proposes a lightweight data report scheme. The scheme uses an online/off-line hash tree-based mechanism and data integrity verification mechanism in order to protect user's privacy and data integrity simultaneously. In addition, the scheme in \cite{90} is efficient in terms of computation cost and communication overhead compared to both the schemes \cite{103,76}.

He et al. in \cite{92} proposes a scheme called AKD that can provide smart meter anonymity and mutual authentication. By adopting Schnorr's signature scheme \cite{100}, the AKD scheme is efficient in terms of computation cost and communication cost compared to the scheme presented in \cite{108}. The AKD scheme is resistant to impersonation attack, replay attack, modification attack, and man-in-the-middle attack, but many assumptions are needed in order to understand the implementation. Ni et al. \cite{93} proposes a scheme called EDAT in order to guarantee the integrity of consumption reports. Based on three main phases, namely, 1) report generation, 2) report aggregation, and 3) report reading, the EDAT scheme is efficient in terms of communication overhead and computational overhead compared to the schemes \cite{109,110,76,111}. Dimitriou et al. \cite{94} proposes two similar protocols, namely, HC and BHC, for privacy of participating meters in order to provide secure aggregate collected measurements. The HC protocol uses random numbers to secure the privacy of the measurements against Honest-but-Curious (HC) behavior while the BHC protocol against Beyond Honest-but-Curious. Tsai et al. in \cite{108} address the anonymity of smart meter by proposing a scheme that achieves mutual authentication and smart meter anonymity at the same time. In addition, the proposed scheme is efficient in terms of computation cost compared to the schemes \cite{112,113,114}.
 \begin{figure}[h]
    \centering
    \includegraphics[width=0.7\linewidth]{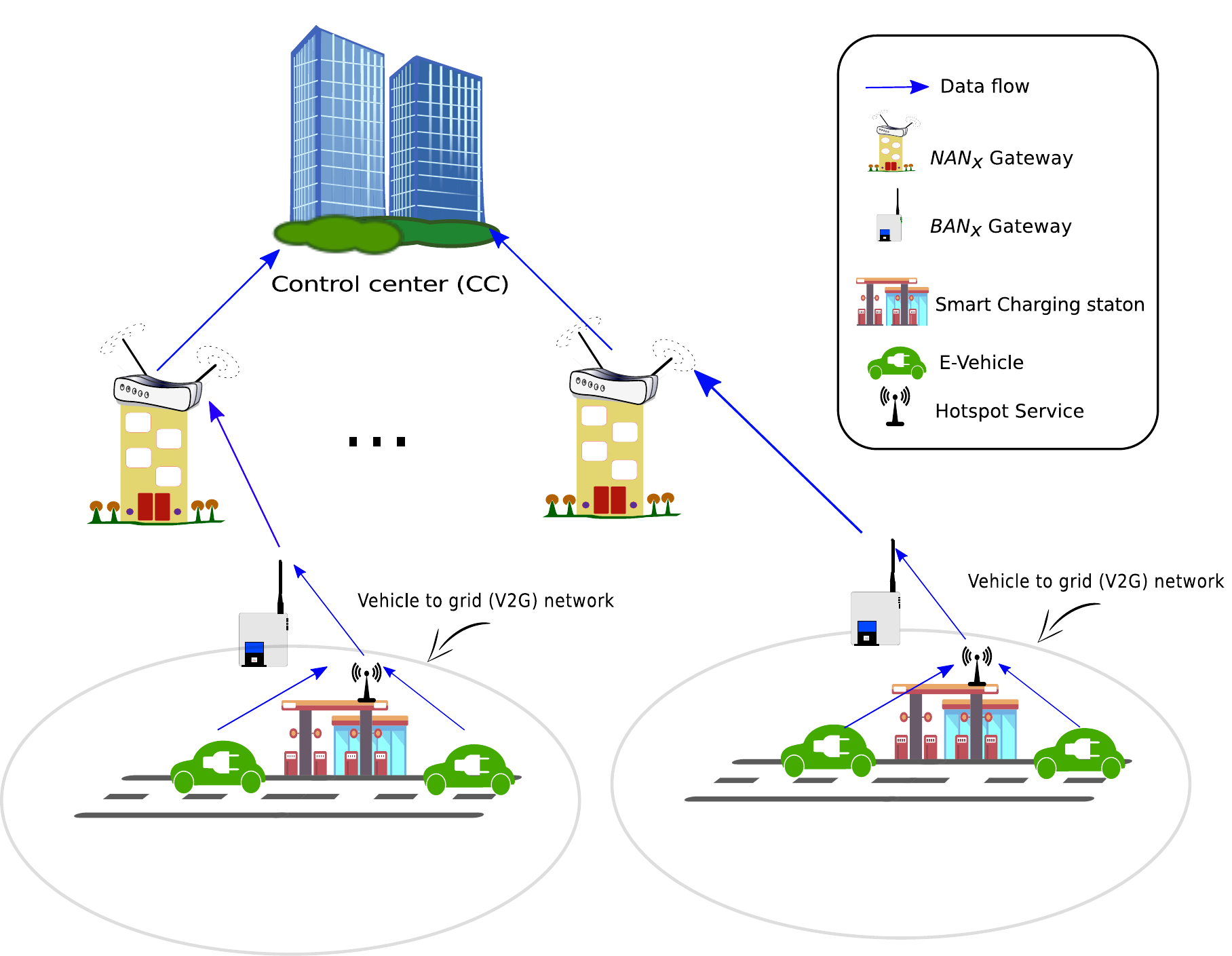}
    \caption{System model used in vehicle-to-grid architecture}
    \label{fig:Fig10}
    \end{figure}
\subsection{Vehicle-to-grid architecture}
The vehicle-to-grid (V2G) is a concept of the integration of the vehicles in Smart Grid as a distributed resource--load and generation/storage device \cite{141}. Fig. \ref{fig:Fig10} shows an example of vehicle-to-grid architecture used in Smart Grid which is based on five parties, namely, e-vehicles, hotspot service in smart charging station, NAN gateway, BAN gateway, and control center. However, as discussed in the recent survey of Han et al. in \cite{25}, V2G network is one of the significant architectures of Smart Grid where electric vehicles (EVs) communicate with service providers via aggregators or other networks. In this subsection, we will review only the works published in 2016 that are not reviewed in the survey \cite{25}. 

Wan et al. in \cite{89} proposes a scheme called PRAC in order to preserve the privacy of electric vehicles owners. Based on four main phases, namely, 1) EV's registration by the trusted third party, 2) EV's anonymous authentication, 3) local aggregators service rewarding to EV, and 4) maintenance for EVs, the PRAC scheme can provide anonymous authentication and rewarding, unlinkable credentials and rewards, a fair exchange for service rewarding, non-repudiation, and efficient system maintenance. Han et al. in \cite{118} proposes a new architecture and practical data management system, called IP2DM, which is based on local aggregators in V2G networks. Specifically, IP2DM can protect the privacy of users during data collection, aggregation, and publication processes. Based on six main techniques involved in IP2DM architecture as onion-level encryption \cite{119} and Homomorphic encryption \cite{54}, the method is robust against many attacks, e.g., replay attack, impersonation attack, chosen ciphertext attack, known plaintext attack, physical attack, and inference attack.

Liu et al. in \cite{124} considers three parties, namely the user, the reservation service provider, and the charging station. Based on reservation and charging processes, the scheme in \cite{124} can provide the privacy-preserving reservation system for charging station, but needs further analysis in order to test the resistance of the method against some attacks such as IP hijacking, distributed denial-of-service attack, and man-in-the-middle attack. In order to resist against these attacks, Saxena et al. in \cite{125} considers three main entities, namely, the electric vehicles, local aggregator, and certification/registration authority. Hence, some privacy models considered in \cite{125} are not considered in the schemes \cite{89,118,124} such as forward privacy, vehicle identity anonymity, and vehicle untraceability. The scheme in \cite{125} can provide not only mutual authentications between the electric vehicles and the local aggregator, and between the electric vehicle and the certification/registration authority, but also can preserve privacy of the electric vehicle's identity, location, charge/discharge selection, expected time, and battery status. In addition, the scheme presented in \cite{125} is efficient in terms of computation and communication overhead compared to the schemes in \cite{130,131}. In another recent work Abdallah et al. in \cite{126} proposes a scheme in order to preserve the privacy of electric vehicles. The scheme in \cite{126} is efficient in terms of communication overhead and computation complexity compared to the scheme \cite{131}.

\begin{table}[h]
\centering
\caption{Future challenges and open directions}
\begin{tabular}{||p{3.4in}||} \hline 
\begin{center}
\underbar{Open directions\newline}
\end{center} \\\begin{itemize}
\item New privacy-preserving schemes against Sybil attacks in Smart Grids
\item New privacy-preserving schemes against Forgery attacks in Smart Grids;
\item Detecting and avoiding Wormhole attacks in Smart Grids;
\item Novel Intrusion Detection mechanisms with low overhead for SCADA systems in Smart Grids;
\item New hybrid cyber range for testing security on ICS/SCADA systems in Smart Grids;
\item New privacy-preserving schemes for IoT-driven Smart Grid;
\item New strategies for location privacy assurance in Smart Grids
\item New privacy metrics;
\item New access control systems for cloud computing services in Smart Grids;
\item New privacy-preserving schemes for Internet of Energy (IoE).
\end{itemize} \\ \hline 
\begin{center}
\underbar{Future challenges\newline}
\end{center} \\
\begin{itemize}
\item Unsolved attacks of leaking privacy;
\item Intrusion Detection mechanisms;
\item Privacy-preserving schemes for IoT-driven Smart Grids;
\item Interdependent privacy for Smart Grids;
\item Access control system for cloud computing services in Smart Grids;
\item Internet of Energy (IoE) and privacy-preserving technologies
\item Ethics and Privacy
\end{itemize} \\ \hline 
\end{tabular}
\label{tab:Tab11}
\end{table}
\begin{figure}[h]
 \centering
 \includegraphics[width=1\linewidth]{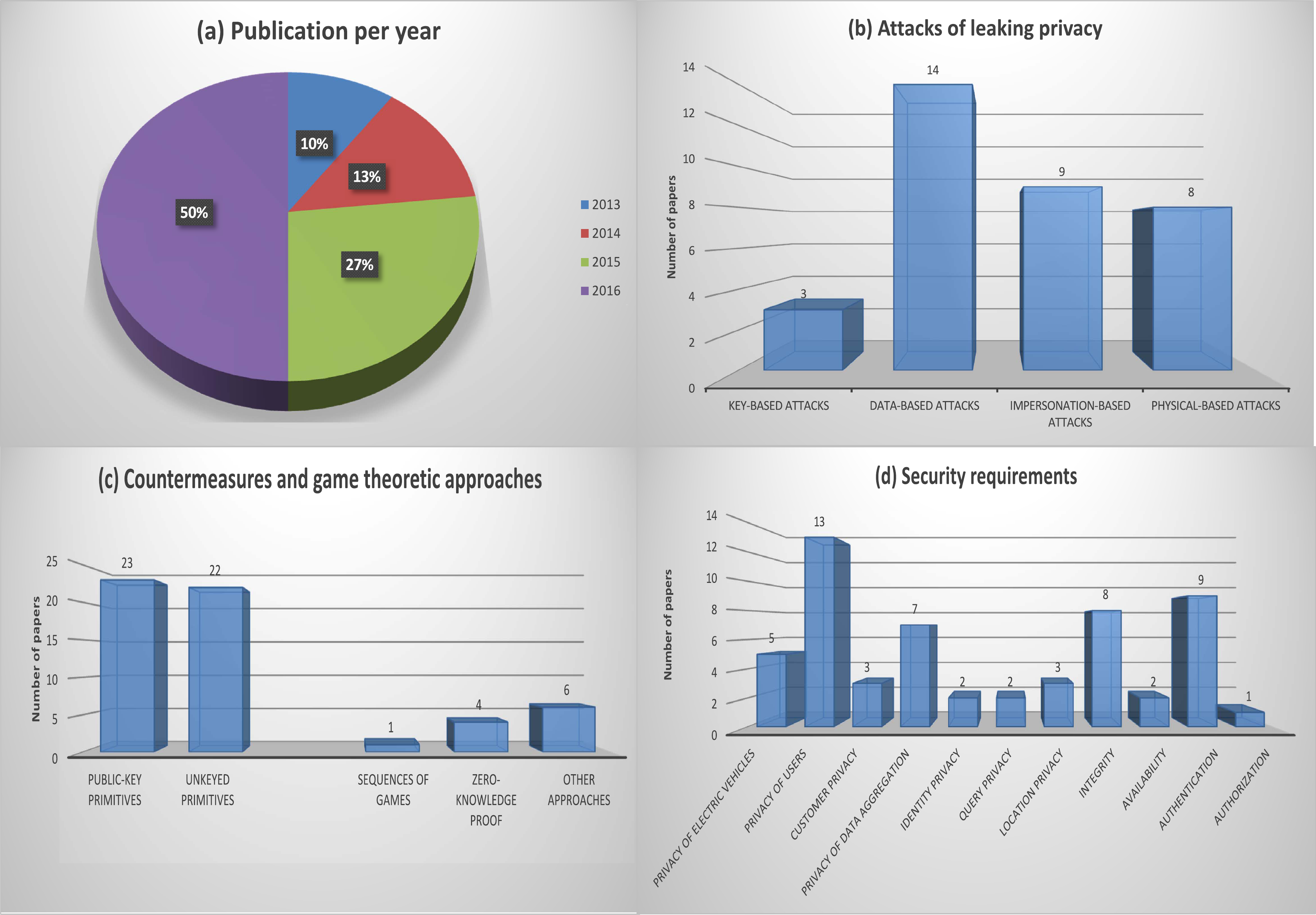}
 \caption{(a) Publication per year, (b) Number of papers vs. Attacks of leaking privacy, (c) Number of papers vs. Countermeasures and game theoretic approaches, and (d) Number of papers vs. Security requirements}
 \label{fig:Fig11}
 \end{figure}
\section{Open questions}\label{sec:open-questions}
As shown in Fig. \ref{fig:Fig11}a, 50\% of the surveyed papers are published in 2016. The privacy-preserving schemes do not encounter sufficiently key-based attacks and they focus on data-based attacks, as shown in Fig. \ref{fig:Fig11}b. The countermeasures used in these schemes use both public-key primitives and symmetric-key primitives, and for game theoretic approaches, they use Zero-Knowledge Proof \cite{86} and other approaches such as strand space model \cite{132}, as shown in Fig. \ref{fig:Fig11}c. In addition, for the security requirements in Smart Grid, these schemes focus on the privacy of users, privacy of electric vehicles, privacy of data aggregation, integrity, and authentication, as shown in Fig. \ref{fig:Fig11}d. To complete our overview of privacy-preserving schemes for Smart Grid communications, technical challenges and open directions for future research are described below and are summarized in Tab. \ref{tab:Tab11}.

\begin{itemize}
\item \textbf{Unsolved attacks and possible solutions:} There are some attacks that are not dealt sufficiently in the surveyed privacy-preserving schemes for Smart Grids such as Sybil attacks, Forgery attacks, and Wormhole attacks. These attacks can pose a lot of privacy problems in wireless networks. As the Smart Grid is based on wireless communications to transmit information, the future works addressing detecting and avoiding these attacks will have an important contribution for the Smart Grid privacy. The idea of signalprints proposed by Liu et al. in \cite{163} can detect Sybil attacks in delay-tolerant networks without requiring trust, which can be applied for Smart Grids. Using the attribute-oriented authentication scheme in Smart Grid, the forgery attacks can be detected and avoided, such as the work in \cite{164}. The privacy-preserving scheme in \cite{165} can detect and avoid wormhole attacks on reactive routing for wireless communications, which can be applied for Smart Grids.

\item \textbf{Intrusion Detection mechanisms for Smart Grids privacy:} Intrusion Detection Systems (IDSs) are not dealt in the surveyed privacy-preserving schemes for Smart Grids. Therefore, IDSs for SCADA system is an open issue that we are working on \cite{139,166,167}. Since the IDS systems can threaten users' privacy \cite{168}, the future works addressing the limitations from both domains will have an important contribution for the Smart Grid privacy. In addition, we believe further research is needed to develop new open access big sets of data to be audited, which include a wide variety of intrusions simulated in a Smart Grid environment.

\item \textbf{Privacy-preserving schemes for IoT-driven Smart Grids:} The need for better privacy is an open issue for the IoT-driven Smart Grids. A work recently published in 2016 by Wu et al. \cite{169} address the mutual privacy and authentic advertisements for the Internet of Things (IoT). Specifically, the private mutual authentication protocol of Wu et al. \cite{169} use  cryptographic primitives such as Identity-based encryption \cite{170} and prefix encryption. Therefore, how to extend this private mutual authentication for the IoT-driven Smart Grids? Hence, privacy, discovery, and authentication for the IoT-driven Smart Grids is one of the future works.

\item \textbf{Interdependent privacy for Smart Grids:} Interdependent privacy for Smart Grids refers to actions of one smart meter affect the privacy of other smart meters. Note that the Interdependent privacy has been defined as a recommendation for further research in our survey on privacy-preserving schemes for ad hoc social networks \cite{81}. As discussed in the survey on technical privacy metrics \cite{171}, there are two options for measuring interdependent privacy that can be used by privacy-preserving schemes for Smart Grids, i.e., measure or create new metrics. Hence, the future works addressing the Interdependent privacy will have an important contribution to improving smart meters privacy in the Smart Grids.

\item \textbf{Access control system for cloud computing services in Smart Grids:} There exist some contributions of cloud technologies developed for Smart Grids, which is surveyed in \cite{175}. The need for access control system is an open issue for the cloud computing services in Smart Grids. A work recently published in 2016 by Liu et al. \cite{174} address the access control system for web-based cloud computing services depends on expressing the attribute predicate as a monotone span program. Therefore, how to extend this system to the cloud computing services in Smart Grids? Hence, privacy and access control for the cloud computing services in Smart Grids is one of the future works.

\item \textbf{Privacy-preserving schemes for Internet of Energy (IoE):} Internet of Energy (IoE) is a new trend in the future of energy where the Internet is interfacing with the power grid, such as the EU project named ARTEMIS-project in \cite{177} that brings together 38 partners from 10 European countries with the total budget Euro 45 million. A recent survey published in 2016 \cite{178} surveys the advances and state-of-the-art technologies of cyber-physical advances for IoE, none of them carries study for the privacy-preserving in IoE. Proposing new privacy-preserving schemes for the IoE is one of the future works.

\item \textbf{Ethics and Privacy:} Users typically expect that the information collected for a specific role and context is treated according to the ethical information norms of the respective communication context. Recently, Spiekermann \cite{spiekermann2015ethical} published a book about ethical IT design where she discusses several aspects of ethics in IT. Based on the material presented in this book, one major issue that arises is how to reassure that all the software running in a smart grid is 'ethical'. 
 
\end{itemize} 

\section{Conclusion}\label{sec:conclusion}
In this article, we surveyed the state-of-the-art of privacy-preserving schemes for Smart Grids, which are published between 2013 and 2016. We presented the survey articles published in the recent years that  describe, namely, Smart Grid communications, Smart Grid applications, Smart Grid security, and Smart Grid privacy. We also presented the major threats of leaking privacy in Smart Grids, including, key-based attacks, data-based attacks, impersonation-based attacks, and physical-based attacks. We reviewed the countermeasures, game theoretic, and formal proof models proposed for Smart Grids used by privacy preserving schemes. We presented a side-by-side comparison in a tabular form for the current state-of-the-art of privacy-preserving schemes (thirty) proposed for Smart Grids. As we have reviewed, privacy-preserving schemes for Smart Grids have advanced significantly in recent years, especially because the need for better privacy for power industry has increased. There is still various very fruitful and challenging research areas (e.g. detecting and avoiding new attacks, IDS architectures for Smart Grid privacy, IoT-driven Smart Grids, new privacy metrics, Privacy for IoE) that need to be further developed in the future.

\bibliographystyle{IEEEtran}
\bibliography{SurveySG}

\end{document}